\documentclass[lettersize,journal]{IEEEtran}
\usepackage{amsmath,amsfonts}
\usepackage{algorithmic}
\usepackage{algorithm}
\usepackage{array}
\usepackage[caption=false,font=normalsize,labelfont=sf,textfont=sf]{subfig}
\usepackage{textcomp}
\usepackage{stfloats}
\usepackage{url}
\usepackage{verbatim}
\usepackage{graphicx}
\usepackage{cite}

\usepackage{kotex}
\PassOptionsToPackage{table,dvipsnames}{xcolor} 
\usepackage{xcolor}
\usepackage{colortbl}
\usepackage[colorlinks=true, allcolors=black]{hyperref}
\usepackage{booktabs}
\usepackage{siunitx}
\usepackage{makecell}
\usepackage{wasysym}
\usepackage{bm}
\usepackage{amssymb}
\newcommand{\system}{Unlearning Comparator}
\newcommand{\secref}[1]{\hyperref[sec:#1]{Section~\ref*{sec:#1}}}
\newcommand{\taskref}[1]{\hyperref[task:#1]{\textbf{#1}}}
\newcommand{\figref}[1]{\hyperref[fig:#1]{Fig.~\ref*{fig:#1}}}
\newcommand{\fignref}[1]{\hyperref[fig:#1]{~\ref*{fig:#1}}}
\newcommand{\equref}[1]{\hyperref[eq:#1]{Eq.~\ref*{eq:#1}}}
\newcommand{\tabref}[1]{\hyperref[tab:#1]{Table~\ref*{tab:#1}}}
\newcommand{\findref}[1]{\hyperref[finding:#1]{\emph{Finding~#1}}}
\newcommand{\findnref}[1]{\hyperref[finding:#1]{\emph{#1}}}

\definecolor{purple}{HTML}{000000}
\definecolor{green}{HTML}{000000}
 
\newcommand{\rev}[1]{#1} 
\newcommand{\del}[1]{}

\begin{document}
\title{\system: A Visual Analytics System for Comparative Evaluation of Machine Unlearning Methods}
\bstctlcite{IEEEexample:BSTcontrol}
\author{Jaeung Lee, Suhyeon Yu, Yurim Jang, Simon S. Woo, and Jaemin Jo
% <-this % stops a space
% \thanks{M. Hayashi is with Graduate School
% of Mathematics, Nagoya University, Nagoya,
% Japan}
% \thanks{M. Owari is with the Faculty of
% Informatics, Shizuoka University,
% Hamamatsu, Shizuoka, Japan.}
\thanks{Jaeung Lee, Yurim Jang, Simon S. Woo, and Jaemin Jo are with Sungkyunkwan University, Suwon, Korea. E-mail: dlwodnd00@skku.edu, \{jyl8755, swoo\}@g.skku.edu, jmjo@skku.edu.}
\thanks{Suhyeon Yu is with Rice University, Houston, TX, USA. E-mail: sy118@rice.edu.}
\thanks{Jaemin Jo is the corresponding author.}
\thanks{Manuscript received xx; xx.}
\thanks{Digital Object Identifier xx.}
}

\markboth{Journal of \LaTeX\ Class Files,~Vol.~xx, No.~xx, xx}%
{Shell \MakeLowercase{\textit{et al.}}: A Sample Article Using IEEEtran.cls for IEEE Journals}


\maketitle

\begin{abstract}
Machine Unlearning (MU) aims to remove target training data from a trained model so that the removed data no longer influences the model's behavior, fulfilling ``right to be forgotten'' obligations under data privacy laws.
Yet, we observe that researchers in this rapidly emerging field face challenges in analyzing and understanding the behavior of different MU methods, especially in terms of three fundamental principles in MU: accuracy, efficiency, and privacy.
Consequently, they often rely on aggregate metrics and ad-hoc evaluations, making it difficult to accurately assess the trade-offs between methods.
To fill this gap, we introduce a visual analytics system, \emph{Unlearning Comparator}, designed to facilitate the systematic evaluation of MU methods.
Our system supports two important tasks in the evaluation process: model comparison and attack simulation. First, it allows the user to compare the behaviors of two models, such as a model generated by a certain method and a retrained baseline, at class-, instance-, and layer-levels to better understand the changes made after unlearning.
Second, our system simulates membership inference attacks (MIAs) to evaluate the privacy of a method, where an attacker attempts to determine whether specific data samples were part of the original training set.
We evaluate our system through a case study visually analyzing prominent MU methods and demonstrate that it helps the user not only understand model behaviors but also gain insights that can inform the improvement of MU methods.
\end{abstract}


\section{Introduction}
\IEEEPARstart{M}{achine} Unlearning (MU) refers to the process of making a trained model ``forget'' specific data it was trained on, such that those data no longer influence the model’s behavior~\cite{cao2015towards}.
This capability has become increasingly important as individuals invoke the ``right to be forgotten''~\cite{rosen2011right} under data privacy laws like the General Data Protection Regulation (GDPR)~\cite{hoofnagle2019european}. 
Regulators have also begun to enforce this: in a 2021 case, the U.S. Federal Trade Commission (FTC) ordered a company to delete not only the unlawfully collected data but also any models derived from it~\cite{singer2021ftc}.
Beyond legal compliance, the ability to reliably remove data is now recognized as a critical component for ensuring the safety and trustworthiness of advanced AI systems~\cite{bengio2025internationalaisafetyreport}.
However, na\"ively retraining a model from scratch for each unlearning request is highly inefficient since modern machine learning models often require a significant amount of time and computational cost to train~\cite{touvron2023llama}.
To address this, MU methods aim to approximate the result of retraining with minimal adjustments to the model's parameters.

We observe that researchers in this emerging field struggle to analyze and understand the behavior of different MU methods.
We identify two main hurdles to this challenge.
First, despite the recent introduction of MU methods~\cite{jia2023model,fan2023salun, cadet2024deep, chen2023boundary, goel2022towards}, there is still no standardized evaluation protocol for systematically comparing them. 
Due to the absence of such a protocol, the MU methods have been separately assessed using disparate metrics~\cite{guo2019certified,becker2022evaluating,chundawat2023zero,neurips-2023-machine-unlearning,liu2025rethinking}.
This not only hinders direct comparisons among methods but also makes it difficult to understand nuanced trade-offs between the methods.
Second, existing evaluations primarily rely on quantitative metrics, providing limited insight into the behaviors of different methods and failing to uncover the underlying decision-making processes happening in the model, which is insufficient to address the analytic tasks we identified. 

% For example, one important task in MU research is to evaluate to what extent an MU model satisfies the \emph{privacy} principle. This principle states that after unlearning, it should be guaranteed that no residual signals of the ``forgotten'' data remain, thereby thwarting attack~\cite{shokri2017membership,fredrikson2015model,chen2021machine,carlini2022membership,hu2024learn}.
\rev{One critical principle in MU evaluation is the \emph{privacy},} which requires that no residual signals of the forgotten data remain, thereby thwarting attacks~\cite{shokri2017membership,fredrikson2015model,chen2021machine,carlini2022membership,hu2024learn}.
For instance, an attacker might query an online medical diagnostic model to check whether a particular patient’s records still shape its responses, which is an example of membership inference attacks (MIAs).
Relying on metrics alone is insufficient for such an evaluation, as such summary metrics would fail to reveal which specific samples remain vulnerable or how model behavior exhibits subtle shifts after unlearning. 
% These limitations call for an interactive approach that allows researchers to simulate attacks and observe how the model responds.
\rev{Consequently, there is a need for an interactive diagnostic approach that enables researchers to verify granular privacy risks through attack simulations, instead of being limited to aggregate metrics.}

To address the two hurdles, we present \emph{Unlearning Comparator}, a visual analytics system that enables systematic and multi-foci comparison between MU methods.
Our system supports the side-by-side comparison between two models at three aspects: \emph{metrics}, \emph{embeddings}, and \emph{attacks}.
By spanning class- and instance-level data as well as output and internal representations, our system aims to show how thoroughly each method removes targeted data without affecting the remainder, highlighting subtle representation changes that aggregate metrics overlook.
\rev{In our case study and interviews, we demonstrate how this systematic comparison enables a human-in-the-loop process, where researchers combine algorithmic insights with visual evidence to integrate ideas from existing methods into a novel MU method design.}
% In our case study and interviews, we demonstrate how this systematic comparison not only clarifies each method’s behavior but also leads to refinements and new designs for a novel MU method.
We contribute:
\begin{itemize}
    \item A design study with MU researchers, incorporating domain analysis to identify user tasks and derive a four-stage workflow;
    \item \emph{Unlearning Comparator}, a visual analytics system that supports the comparative evaluation of MU methods through visualizations of metrics, embeddings, and simulated attacks;
    \item The first (to our knowledge) visual analysis of prominent MU methods, which uncovers overlooked behavioral patterns and informs the development of a novel, more effective MU method, \emph{Guided Unlearning (GU).}
\end{itemize}
% \vspace{1mm}
\section{Background and Related Work}
\subsection{Machine Unlearning Fundamentals}
\label{sec:metric}
We call the model trained on the entire dataset before unlearning the \emph{original} model.
As the input to an MU method, the user chooses the data to be unlearned by partitioning the entire dataset into a \emph{retain set} (data to keep) and a \emph{forget set} (data to remove).
The goal of MU is to produce an \emph{unlearned} model by efficiently modifying the original model's parameters to approximate the behavior it would exhibit if the forget set had never been included.
The performance and behavior of the unlearned model are often compared with a \emph{retrained} model, which is obtained by training the model from scratch on the retain set.
While building a retrained model is inefficient in practice, it serves as a gold standard to assess the effectiveness of MU methods.

The MU methods are commonly evaluated based on three principles: accuracy, efficiency, and privacy~\cite{xu2024machine, nguyen2022survey, zhang2023review, sai2024machine}.

\textbf{Accuracy.}
The accuracy principle requires that the forget set should be unlearned without significantly degrading the model’s performance on the retain set. 
Since deep neural networks have intertwined internal representations, simply dropping certain parts of the representations may harm the overall accuracy.
We use four standard accuracy measures: \emph{Unlearning Accuracy (UA)} measures correctness on the forget set (lower is better), while \emph{Retain Accuracy (RA)} measures correctness on the retain set (higher is better).
If a separate test set is available, the accuracy can also be measured for the unseen data, resulting in \emph{Test Unlearning Accuracy (TUA)} (lower is better) and \emph{Test Retain Accuracy (TRA)} (higher is better).

\textbf{Efficiency.} The efficiency principle relates to the computational overhead that an MU method exhibits; for example, building a retrained model can be considered as an MU method but is inefficient as it requires full retraining on the retain set. 
The efficiency of a method is quantified by its \emph{Run Time (RT)}, measured in seconds.

\textbf{Privacy.}
The privacy principle ensures the true removal of the forget set without lingering signals to prevent attackers from detecting data influence. In practice, this is primarily evaluated through the success rate of membership inference attacks (MIAs)~\cite{shokri2017membership,chen2021machine,carlini2022membership}, which predicts whether specific data samples were part of the training set based on the model's output.
However, this single metric is often insufficient, as it fails to reveal which specific samples remain vulnerable and why, thus requiring more nuanced, interactive analysis.

\subsection{Machine Unlearning Methods}
\label{sec:method}
To avoid the computational overhead of the retraining, most MU methods take approximate approaches that adjust the model's weights without completely rebuilding it.
We surveyed the literature and identified three widely adopted MU methods in the context of approximate unlearning. These baselines appear across many recent MU studies~\cite{cadet2024deep, golatkar2020eternal} and represent complementary strategies.

\begin{itemize}
    \item \textbf{Fine-Tuning (FT)}: This method continues training (i.e., running more training epochs) only on the retain set, relying on catastrophic forgetting~\cite{goodfellow2013empirical} of the forget set.
    \item \textbf{Random Labeling (RL)}: This method assigns random labels to the forget set and fine-tunes the model on this modified dataset, causing the model to forget the original information associated with the forget set.
    \item \textbf{Gradient Ascent (GA)}: This method adjusts the model's parameters to maximize loss on the forget set to intentionally unlearn it.
\end{itemize}

Beyond these baselines, more advanced methods have also been proposed recently. 
For example, \textbf{SCRUB}~\cite{kurmanji2023towards}, which uses a teacher–student distillation framework to maximize loss on the forget set while minimizing loss on the retain set, and \textbf{Saliency Unlearning (SalUn)}~\cite{fan2023salun}, which identifies and masks weights most influenced by the forget set before applying random labeling plus targeted fine-tuning. 
Although these methods represent diverse strategies, few studies compared them under a consistent protocol.
Our system addresses this gap by allowing researchers to examine and contrast their trade-offs in accuracy, efficiency, and privacy.

\subsection{Distribution-based Approach for Privacy Evaluation}
\label{sec:dist}
While accuracy and efficiency are relatively straightforward to assess, privacy evaluation is more challenging. 
In classification tasks, attackers often employ prediction entropy (i.e., the spread of class probability outputs) as a signal for membership inference. In this entropy-based MIA (E-MIA), a sample is predicted as ``non-training'' if its entropy exceeds a chosen threshold (i.e., the model is uncertain about the sample), and ``training'' otherwise (i.e., the model is certain).

However, evaluating a model’s privacy by simulating such attacks may be less reliable in unlearning scenarios, as these attacks can be easily circumvented; for example, one can deliberately make a model flatten its output scores to increase the entropy and thereby appear resilient to E-MIA.
To provide a more rigorous privacy evaluation, we adopt and extend a stronger MIA from prior work~\cite{neurips-2023-machine-unlearning}, which assumes that the attacker compares the unlearned and retrained models’ output distributions. 
If the unlearned model truly behaves as if it never saw the forget set, its distribution should align with that of the retrained model, providing stronger assurance that no residual data signals remain; in our extension, we incorporate both confidence- and entropy-based attacks under multiple thresholds, yielding a unified \emph{Worst-Case Privacy Score} (see~\secref{PS}).

\subsection{\rev{Visualization for Model Understanding and Comparison}}
Visual analytics has been widely employed to help people develop, interpret, \text{compare}, and improve machine learning models~\cite{subramonyam2023we, hohman2018visual, wang2024visual}.
% It aims to bridge the gap between complex model behaviors and human understanding, which is essential for building trustworthiness and reliability~\cite{chat2024trust}.
% For example, Uni-Evaluator~\cite{chen2023unified} supports model evaluation through unified performance analysis across multiple tasks to guide debugging. 
% To help understand internal mechanisms of models, interactive systems that visualize layer-wise activations~\cite{kahng2017cti} or architecture‑specific explanations for CNNs~\cite{wang2020cnn} and Transformers~\cite{yeh2023attentionviz, cho2025transformer} have also been studied. 
\rev{To bridge the gap between complex model behaviors and human understanding~\cite{chat2024trust}, prior work has introduced systems ranging from unified performance evaluation~\cite{chen2023unified} to internal mechanism inspection via layer-wise activations~\cite{kahng2017cti} or architecture-specific explanations for CNNs~\cite{wang2020cnn} and Transformers~\cite{yeh2023attentionviz, cho2025transformer}.}
Furthermore, attribution techniques such as Grad-CAM~\cite{selvaraju2017grad} are now widely used to highlight important regions in input images.
While these systems and techniques help analyze a single model, they often fall short when assessing differences between models in a comparative manner, which is also common in machine learning.

\rev{Existing visual analytics systems for model comparison provide workflows for revealing relative strengths, weaknesses, and behavioral differences between models.
For example, Manifold~\cite{zhang2018manifold} supports pairwise comparison to localize performance gaps, and ModelWise~\cite{meng2022modelwise} enables multi-model comparison to surface when and why models disagree for iterative improvement.
Other domain-specific systems compare models for compression trade-offs~\cite{boggust2024compress}, architectural differences of CNNs~\cite{xuan2022vac}, or behavioral shifts in language model adaptation~\cite{sevastjanova2022visual}.
However, these generic workflows are inadequate for the unique design space of MU.
While comparison typically aims to identify a ``better'' model under fixed ground-truth labels, MU requires behavioral verification: assessing whether an unlearned model aligns with a ``gold-standard'' retrained model that varies with the forget set.
Moreover, the unlearning process involves opposing optimization objectives: removing information in the forget set while preserving it in the retain set.
Systems treating datasets holistically cannot effectively guide the trade-off analysis required for this dichotomy.
Finally, privacy in MU is a behavioral property assessed through adversarial verification (e.g., MIAs), an aspect outside standard model comparison workflows.
}

To address this distinct design space, \rev{we present a visual analytics system that formalizes these elements into a unified workflow} for the comparative evaluation of MU methods, thereby advancing research in this emerging field.

% Model comparison aims to reveal relative strengths, weaknesses, and behavioral differences between models.
% Several visual analytics systems have been developed to support comparisons in certain domains of machine learning.
% Examples include analyzing efficiency-behavior trade-offs in model compression~\cite{boggust2024compress}, comparing the internal structures of CNN architectures~\cite{xuan2022vac}, and evaluating behavioral shifts when adapting language models to new tasks~\cite{sevastjanova2022visual}. 
% However, these comparison tools are not designed for MU, a nascent field with unique, domain-specific challenges: new concepts such as the distinction of forget and retain sets, the notion of a retrained model as a gold standard, and the need for privacy evaluation.

% Overall, although prior VIS4ML systems have effectively supported model understanding across domains, MU introduces distinctive challenges for visual analytics.
% To address this gap, we present a visual analytics system that enables a unified workflow for the comparative evaluation of MU methods, thereby advancing research in this emerging field.
\section{Domain Analysis}
\label{sec:domain}
Our target users are MU researchers who want to compare the performance of different MU methods in terms of the three principles and understand their trade-offs.
To identify their needs, we conducted a design study by holding \rev{weekly} discussions over two months with two MU experts \rev{(one professor specializing in ML security including the privacy concerns of MU and one graduate student with years of experience in MU research)} to gather insights into their analytical challenges and integrating their feedback throughout the development process. 
We also surveyed prior literature, including recent benchmarking studies~\cite{cadet2024deep, jia2023model, goel2022towards}, to extract common analysis and evaluation patterns.
Synthesizing the insights from both sources allowed us to summarize their goals as follows: 
(1) evaluate how different MU methods unlearn data by enabling pairwise comparisons as a unified workflow, and (2) leverage these insights to improve MU methods aligned with the three principles.

While MU is an active research area in various domains such as image classification, text generation, and image generation, we focus specifically on class-wise unlearning for image classification as it provides the most widely used and robust environment for systematically evaluating MU methods.
From this point, for simplicity, we refer to the ``forget set'' and ``retain set'' as the ``forget class'' and ``retain classes,'' respectively.

\subsection{User Tasks}
Based on the insights gained from the design study, we distilled the following five user tasks that our system must support:

\textbf{T1. Build and Screen Models.}
\label{task:T1}
Many MU methods lack standardized training recipes as their optimization objectives and resulting loss scales can be unpredictable compared to standard training. This forces experts to experiment extensively with different MU methods and their hyperparameters. Therefore, an initial step is to build multiple candidate models and collect their summarized performance metrics. This allows experts to efficiently screen candidates and make an informed selection of a pair of models for a deeper comparison.

\textbf{T2. Compare Two Models Pairwise.}
\label{task:T2}
We found that the experts compare two models to understand how MU has altered a model's behavior.
For example, they compare an unlearned model with a retrained model for benchmarking purposes (see \taskref{T2.2}).
Thus, we focus on supporting these targeted pairwise comparisons. We identify three types of such tasks, each serving a distinct purpose:

\emph{T2.1 Compare the Original Model with the Retrained or an Unlearned Model.}
\label{task:T2.1}
Comparing the original model with the retrained illustrates the ideal scenario in which the forget class was never included, providing a reference for evaluating the MU methods. 
Comparing the original model with an unlearned then reveals how the method has altered the model’s behavior.

\emph{T2.2 Compare the Retrained Model with an Unlearned Model.}
\label{task:T2.2}
This task reveals how closely the unlearned model approximates the gold standard retrained model, which is the goal of MU.

\emph{T2.3 Compare Unlearned Models with Each Other.}
\label{task:T2.3}
Comparing two unlearned models allows experts to assess relative strengths and weaknesses across methods; for example, by comparing a baseline MU method with its optimized variant, they can identify which performs better.

\textbf{T3. Investigate Class-wise Accuracy and Confidence.}
\label{task:T3}
Our experts wanted to investigate if the accuracy for the forget class decreases as desired, without harming the retain classes, which is the accuracy principle.
They also wanted to investigate the confidence (i.e., the probabilities assigned to classes) to reveal subtle behavioral changes missed by the final prediction alone.
However, since distribution shifts due to unlearning can make absolute softmax-based confidence scores misleading, this task focuses on examining patterns of over- and under-confidence, as well as overall calibration quality, rather than absolute confidence scores.

\textbf{T4. Analyze Changes in Layer Representations.}
\label{task:T4}
Even if accuracy metrics indicate successful forgetting, prior studies~\cite{jeon2025idi, seo2025da} have shown that intermediate layers might still encode knowledge of the forget class, compromising the privacy principle.
Hence, analyzing layer activations in the feature space before and after unlearning clarifies how the model’s internal representation changes and whether the forget class has been removed at the layer level.

\textbf{T5. Verify Privacy through Attack Simulation.}
\label{task:T5}
One important task in MU research is to ensure the privacy of the models after unlearning.
To this end, the experts wanted to simulate membership inference attacks (MIAs) on an unlearned model and its retrained counterpart to see if an attacker could still predict whether a specific data point was part of the original training set.

\begin{figure*}[t]
    \centering
    \includegraphics[width=\textwidth]{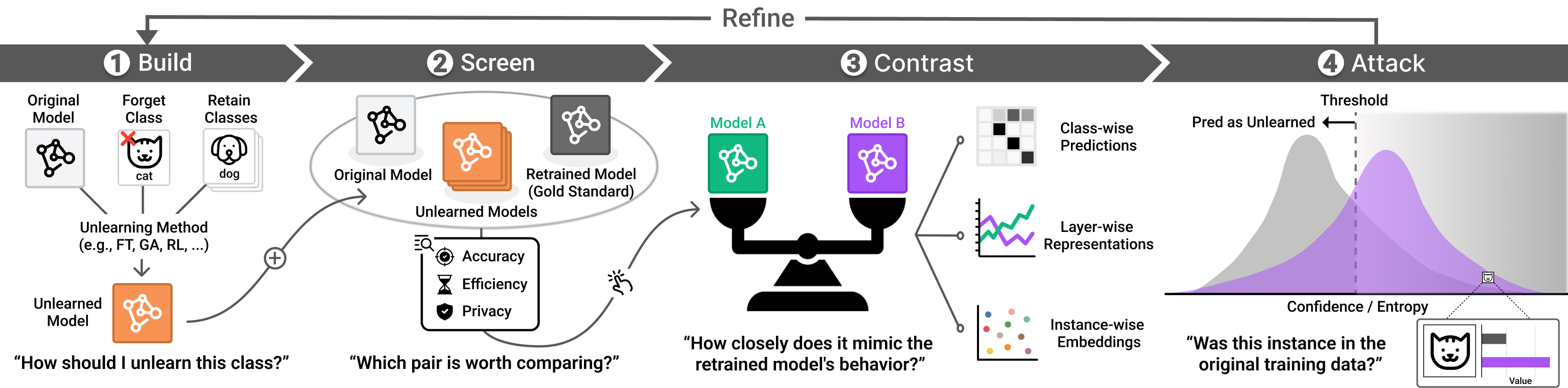}
    \caption{Comparative MU evaluation workflow of \emph{Unlearning Comparator}. The workflow guides users through four iterative stages. In the \emph{Build} stage, users generate various unlearned models, followed by the \emph{Screen} stage where they use summary metrics to select two for in-depth analysis. The \emph{Contrast} stage involves comparing the selected pair from class-, instance-, and layer-level perspectives to understand model behaviors. Finally, the \emph{Attack} stage verifies privacy by simulating membership inference attacks. Insights from all stages then guide iterative refinement of the unlearning methods.}
    \label{fig:workflow}
\end{figure*}

\subsection{Comparative Machine Unlearning Evaluation Workflow} \label{sec:workflow}
We distill the user tasks into a unified process, proposing the \emph{Comparative Machine Unlearning Evaluation Workflow}, which consists of four stages: \emph{Build}, \emph{Screen}, \emph{Contrast}, and \emph{Attack} (see \figref{workflow}).

In the \emph{Build} stage, users specify the base model (e.g., the original model) as a starting point, define the forget class, and choose an unlearning method along with hyperparameters, producing multiple candidate models (\taskref{T1}). 
Next, in the \emph{Screen} stage, they review summarized performance metrics (\taskref{T1}) to decide which models warrant closer inspection.
Users then proceed to the \emph{Contrast} stage, where they compare two chosen models (\taskref{T2}). 
For instance, they can compare an unlearned model with the original model to inspect model changes (\taskref{T2.1}) or with the retrained model to assess proximity to the ideal outcome (\taskref{T2.2}), or compare two unlearned models to contrast their behaviors (\taskref{T2.3}).
They examine accuracy shifts and prediction patterns to check that the forget class is removed without adversely affecting the retain classes (\taskref{T3}). 
They also investigate representation changes across the network and observe the embedding space to see how features have changed after unlearning (\taskref{T4}).
Finally, in the \emph{Attack} stage, users conduct MIAs (\taskref{T5}) to confirm that the forget class is no longer identifiable, exploring different attack metrics and strategies to assess privacy risks and detect vulnerable samples.

After reviewing the insights from these stages, users can refine their method by adjusting hyperparameters or modifying the unlearning algorithm itself, and return to the build stage to iteratively improve accuracy, efficiency, and privacy.
\section{The Unlearning Comparator System}
\begin{figure*}[t]
\centering
\includegraphics[width=\linewidth]{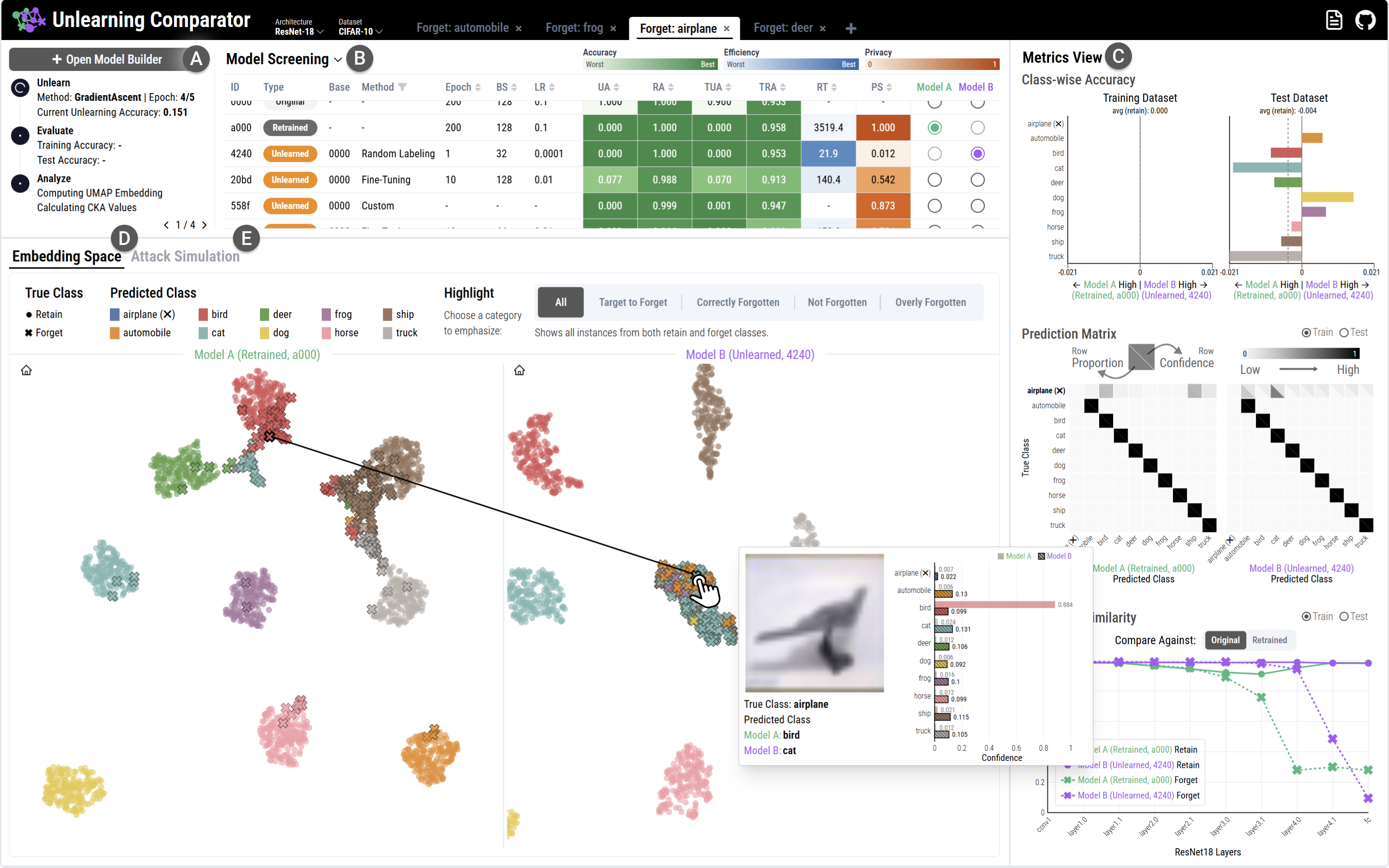}
\caption{
  \emph{\system} assists MU researchers in analyzing and comparing MU methods.
  (A) The \textbf{Model Builder} (shown in \figref{build_screen}A) creates unlearned models.
  (B) The \textbf{Model Screening} view lets users obtain an overview and select two models for deeper inspection.
  (C) The \textbf{Metrics} view highlights class-level performance and internal representation changes.
  (D) The \textbf{Embedding Space} view displays each model’s feature embeddings side-by-side.
  (E) The \textbf{Attack Simulation} view (shown in \figref{attack}) performs membership inference attacks to verify that no residual signal remains about the unlearned data.
}
\label{fig:teaser}
\vspace{1mm}
\end{figure*}

In this section, we elaborate on \emph{Unlearning Comparator}, a visual analytics system designed for comparative evaluation of MU methods, supporting \taskref{T1}-\taskref{T5}.
We chose two datasets, CIFAR-10 \cite{krizhevsky2009learning} and Fashion-MNIST \cite{xiao2017fashion}, to demonstrate our system, as they are commonly employed in prior MU research.
For model architectures, we included two representative architectures: a well-known convolutional architecture ResNet-18 \cite{he2016deep} and a more advanced transformer-based architecture, ViT-B/16 \cite{dosovitskiy2020image}, pre-trained on ImageNet \cite{deng2009imagenet}.
While these choices ensure consistency with existing literature, \rev{our current system targets ten-class image classification datasets and corresponding model architectures. We discuss extensions beyond this setting in~\secref{lim}.}
% While these choices ensure consistency with existing literature, our system is designed to support using other datasets and architectures. 

\subsection{Designing the Worst-Case Privacy Score}
\label{sec:PS}
To facilitate comparative privacy evaluation (\taskref{T5}), we define a \emph{Worst-Case Privacy Score (WCPS)} by synthesizing established MIA concepts. 
There are two widely adopted standard MIAs: (1) Confidence-based MIA (C-MIA)~\cite{yeom2018privacy,jia2023model} and (2) Entropy-based MIA (E-MIA)~\cite{salem2018ml,chundawat2023can}.
Inspired by the NeurIPS 2023 Machine Unlearning Competition~\cite{neurips-2023-machine-unlearning}, we design \emph{WCPS} under the assumption of a stronger, more knowledgeable attacker, as those standard MIAs can be easily circumvented (\secref{dist}).
Under this assumption, the adversary can access both the predicted label’s confidence and the entropy of all logit outputs from unlearned or retrained models. Although this setting is infeasible in practice, it is adopted to consider an optimal adversary and thus facilitate conservative privacy evaluation.
The attacker measures how distinguishable the output distributions (confidence or entropy) of the models are on the forget class to distinguish between ``training'' and ``non-training'' inputs.
A threshold-based decision rule is applied to perform the inference.
We aggregate the results over multiple thresholds and both output statistics (confidence and entropy) to derive a single worst-case score, denoted as \emph{WCPS}.

\textbf{Score Definition.}
Let \( z = (z_1, z_2, \dots, z_n) \) denote the logit outputs for \( n \) classes. The softmax function is given by:
\begin{equation}
p_i = \frac{e^{z_i}}{\sum_{j=1}^{n} e^{z_j}}.
\label{eq:softmax}
\end{equation}

The confidence function \(C(p)\) using the log-odds transformation is defined as:
\begin{equation}
p_{\text{max}} = \max_i p_i,\quad
C(p) = \log(p_{\text{max}}) - \log(1 - p_{\text{max}}).
\label{eq:confidence}
\end{equation}

The entropy \(H(p)\) is defined with temperature \( T = 2.0 \), \rev{chosen empirically to better spread the entropy values for visual inspection, as follows:}
%selected to better separate model output distributions, as follows:
\begin{equation}
p_i^{(T)} = \frac{e^{z_i/T}}{\sum_{j=1}^{n} e^{z_j/T}}, \quad
H(p) = -\sum_{i=1}^{n} p_i^{(T)} \log(p_i^{(T)}).
\label{eq:entropy}
\end{equation}

For each attack, to distinguish whether a sample comes from the retrained or unlearned model, we define threshold-based decision rules over these metrics. These rules produce false positive rates (FPR) and false negative rates (FNR). We consider 100 evenly spaced thresholds across each metric's range, testing both directions of inequality for optimal attack performance.

At each threshold \(t\), let \(\text{FPR}_t\) and \(\text{FNR}_t\) respectively be the false positive and false negative rates observed at that threshold. Then we calculate \(\epsilon_t\) as follows:
\begin{equation}
\epsilon_t = 
\max\!\Bigl(0,\;\min\Bigl(
\log\!\bigl(\tfrac{1 - \text{FPR}_t}{\text{FNR}_t}\bigr),\;
\log\!\bigl(\tfrac{1 - \text{FNR}_t}{\text{FPR}_t}\bigr)
\Bigr)\Bigr).
\label{eq:epsilon}
\end{equation}

Intuitively, \(\epsilon_t\) measures the indistinguishability of the output distributions from the retrained and unlearned models at threshold \(t\), capturing the degree of privacy achieved. A higher \(\epsilon_t\) indicates a more successful attack and thus lower privacy.

Accordingly, at each threshold \(t\), we define the threshold-level Attack Score \((AS_t)\) and Privacy Score \((PS_t)\) as follows:
\begin{equation}
\label{eq:AS-FQS-t}
AS_t = 1 - 2^{-\epsilon_t}, 
\quad
PS_t = 1 - AS_t.
\end{equation}

To obtain a privacy score for each output statistic, we select the minimal \(PS_t\) across all thresholds:
\begin{equation}
PS_{\text{C}} = \min_{t \in T}(PS_t^C), 
\quad
PS_{\text{E}} = \min_{t \in T}(PS_t^H),
\label{eq:PS-C-E}
\end{equation}
where \(PS_t^C\) is computed using the confidence \(C(p)\) from \equref{confidence} at threshold \(t\), and \(PS_t^H\) uses the entropy \(H(p)\) from \equref{entropy}.

Finally, for the overall model, we apply a worst-case perspective across the two output statistics by taking the minimum of their PS values:
\begin{equation}
\label{eq:FQS-model}
WCPS = \min\Bigl(PS_{\text{C}},\,PS_{\text{E}}\Bigr).
\vspace{1mm}
\end{equation}

\textbf{Comparison with Standard MIAs.}
\begin{figure*}[t]
\centering
\includegraphics[width=0.99\textwidth]{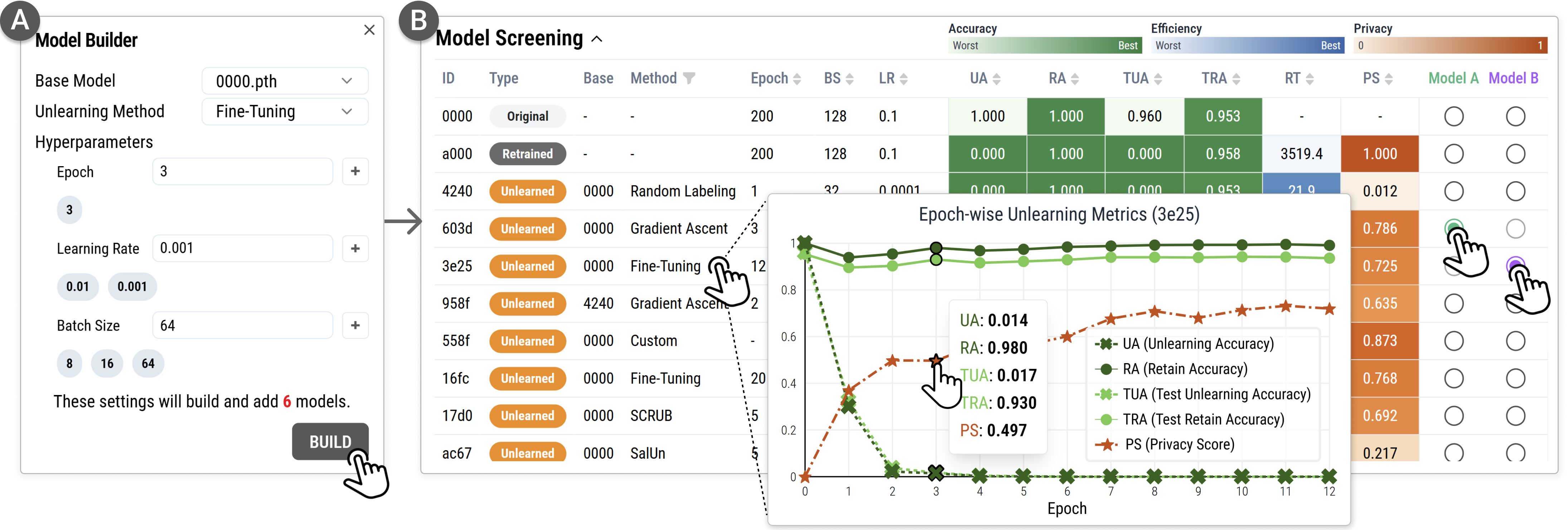}
\caption{
Users configure unlearning settings in the \textbf{Model Builder} (A) to generate candidate models. By selecting multiple values for each hyperparameter, they can build multiple models at once; all combinations are generated automatically.
They then review each model’s performance in the \textbf{Model Screening} view (B), which presents summary metrics and reveals epoch-wise metrics upon clicking a row, allowing users to select two for deeper comparison.}
\label{fig:build_screen}
\vspace{-1mm}
\end{figure*}
\begin{figure}[t]
    \centering
    \includegraphics[width=\linewidth]{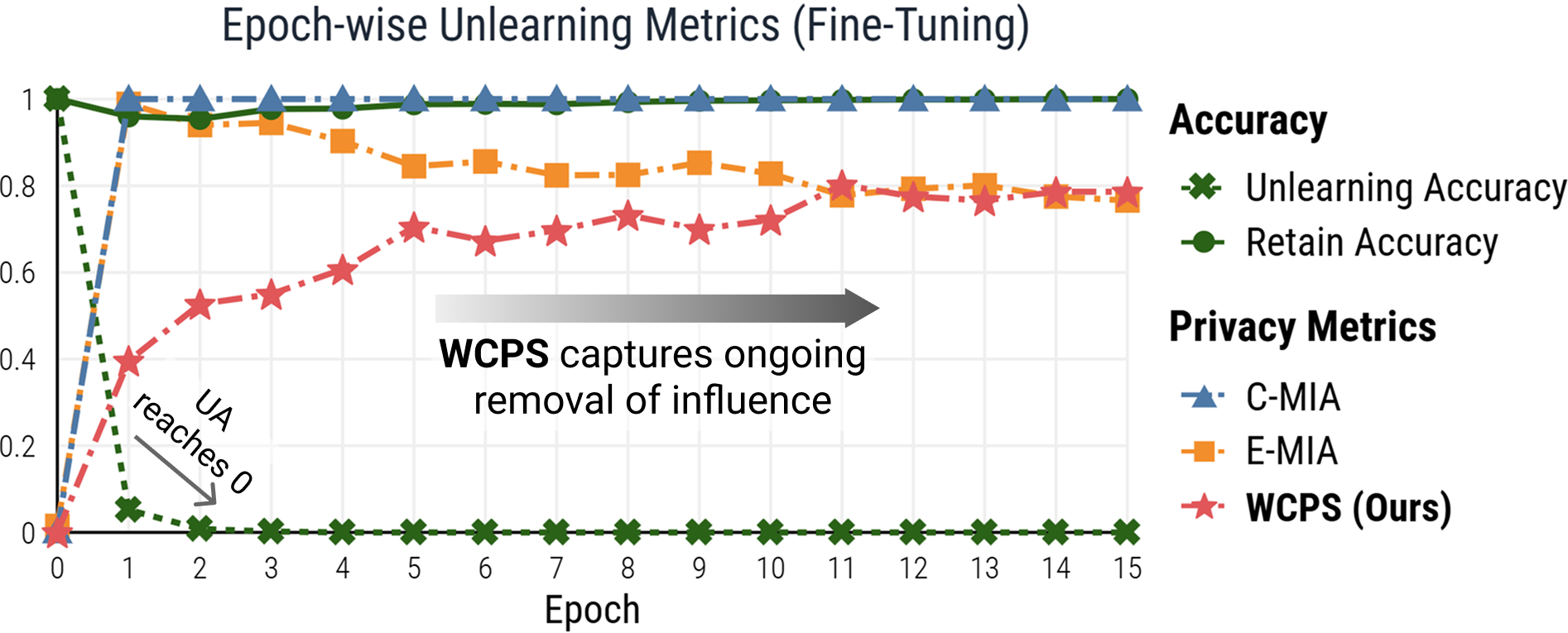}
    \caption{
    Comparison of privacy metrics using FT beyond zero unlearning accuracy. 
    C-MIA prematurely reaches and remains at 1.0 due to its reliance on raw confidence, while E-MIA incorrectly declines as the model confidently misclassifies samples. 
    In contrast, \emph{WCPS} progressively increases, reflecting the convergence toward the retrained model's distribution.
    }
    \label{fig:ps}
    \vspace{-1mm}
\end{figure}
By design, \textit{WCPS} assumes a stronger attacker, providing a more conservative privacy evaluation than standard C-MIA and E-MIA.
To examine the correlation between these metrics in practice, we conducted an experiment using the FT method that simulates the process where the forget class representations are progressively overwritten by the retain classes’ representations.
This continues even after the unlearning accuracy reaches zero, as merely achieving zero accuracy on the forget class does not guarantee complete removal of its influence.
In this comparison, C-MIA and E-MIA are calculated as the proportion of the forget class identified as non-members by the attack.

As shown in \figref{ps}, \emph{WCPS} gradually increases as epochs progress, reflecting that the forget class representations are progressively replaced, and the output distribution converges toward the retrained model's distribution.
However, C-MIA and E-MIA fail to capture this progression. 
Even at epoch 1, where the distribution has not yet converged (\emph{WCPS} $\approx$ 0.4), C-MIA prematurely reaches 1.0 (i.e., classifying all forget class samples as non-members). 
It remains at 1.0 due to its reliance on raw confidence values without log-odds transformation (\equref{confidence}), which would better separate confidence distributions; this reliance causes incorrect membership inference when confidence drops.
E-MIA exhibits even more misleading behavior; it declines as epochs progress because the low entropy from confident misclassification to other classes is misinterpreted as evidence of membership.
Notably, the \emph{WCPS} plateaus around 0.8 rather than reaching 1.0 because FT removes information spanned by the retain classes, struggling to fully erase residual information unique to the forget class~\cite{ding2025ft}; \rev{this motivates stronger or complementary strategies beyond FT to achieve more complete erasure.}

\subsection{User Interface}
\emph{\system}'s user interface (\figref{teaser}) is designed to support each stage of our workflow (\figref{workflow}). As shown at the top of \figref{teaser}, the interface consists of multiple tabs with each providing a dedicated workspace for a specific forget class. A plus icon allows users to add new tabs for additional forget classes.
Each tab consists of the \textbf{Model Builder} (\figref{teaser}A) for the \textit{Build} stage, the \textbf{Model Screening} view (\figref{teaser}B) for the \textit{Screen} stage, the \textbf{Metrics} view (\figref{teaser}C) and the \textbf{Embedding Space} view (\figref{teaser}D) for the \textit{Contrast} stage, and the \textbf{Attack Simulation} view (\figref{teaser}E) for the \textit{Attack} stage.

\subsubsection{Model Building (the \textit{Build} stage, \taskref{T1})}
\label{sec:build}
The \textbf{Model Builder} creates unlearned models (\figref{build_screen}A).
Users specify the base model to unlearn, the chosen method, and multiple hyperparameter combinations. Models corresponding to all these combinations \rev{(epochs $\times$ learning rates $\times$ batch sizes)} are automatically generated.
To avoid complexity while allowing meaningful adjustments, we limit tunable hyperparameters to epochs, learning rate, and batch size, as these have been shown to play a critical role in prior MU studies~\cite{cadet2024deep,kurmanji2023towards} and offer sufficient control over the unlearning process.
We offer three baseline MU methods from~\secref{method} (i.e., fine-tuning, random labeling, and gradient ascent).

To minimize the computational overhead of training new models, we provide both the original model and the class-wise retrained models in advance.
Although such retrained models are rarely available in real-world applications, we include them here as essential benchmarks to enable controlled evaluation of MU methods in a research context.
These models follow standardized benchmarking recipes~\cite{moreau2022benchopt, torchvision2016} to ensure reproducibility and robustness.
In addition to these baselines, users can build their custom models by (1) extending our provided Python code template to implement their own method or (2) uploading weight files for their unlearned models.

The building process is shown in the \textbf{Model Builder} in real time (\figref{teaser}A). Once completed, a newly created model automatically appears in the \textbf{Model Screening} view. 

\subsubsection{Model Screening (the \textit{Screen} stage, \taskref{T1})}
The \textbf{Model Screening} view provides an overview of all models, including both system-provided models (original and retrained) and user-created unlearned models (\figref{build_screen}B). 
Each row in the table shows the model’s basic configuration and representative metrics for the three main MU principles described in \secref{metric}. 
Distinct color schemes are used for these principles: UA, RA, TUA, and TRA for accuracy, RT for efficiency, and the \emph{WCPS} (defined in \secref{PS}) for privacy (abbreviated as PS in the UI for simplicity). 
Users can filter by method or sort the metrics to identify promising models.
Additionally, clicking on any row reveals the per-epoch performance metrics as a line chart (pop-up in \figref{build_screen}B) to show the model's unlearning progress and convergence. After this initial screening, they can select a pair using the two radio buttons in the rightmost columns for closer examination in the subsequent view.

In the pairwise visualization analysis, the selected \textcolor{green}{Model A} and \textcolor{purple}{Model B} are shown using two distinct \textcolor{green}{green} and \textcolor{purple}{purple} colors,  respectively, while the retain class and forget class are shown as symbols, ``\ensuremath{\scalebox{0.75}{$\CIRCLE$}}'' for the retain classes and ``\ensuremath{\textbf{\texttimes}}'' for the forget class. For cases that require color-coding all 10 classes, we apply the Tableau10 color scheme.

\subsubsection{Model Comparison (the \textit{Contrast} stage, \taskref{T2}--\taskref{T4})}
\begin{figure*}[t]
\centering
\includegraphics[width=\textwidth]{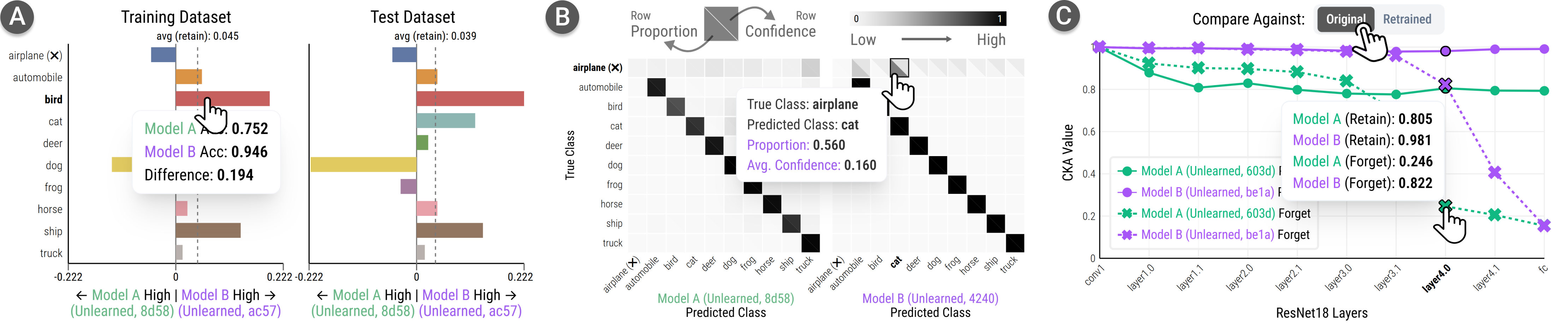}
\caption{
The \textbf{Metrics} view provides metrics that reveal how an unlearning method targets the forget class while preserving the retain classes.
(A) \emph{Class-wise Accuracy} chart displays the per-class accuracy differences to examine high-level trade-offs.
(B) \emph{Prediction Matrix} visualizes predicted proportion and average confidence to inspect misclassification patterns,
and (C) \emph{Layer-wise Similarity} chart shows the similarity of layer representations against the original or the retrained models to reveal changes in internal representations.}
\label{fig:metrics}
\vspace{-2mm}
\end{figure*}

The \textbf{Metrics} view provides metrics that reveal how an MU method targets the forget class while preserving the retain classes.
It comprises three components: (1) \emph{Class-wise Accuracy} chart to analyze accuracy difference for each class, (2) \emph{Prediction Matrix} to capture misclassification patterns, and (3) \emph{Layer-wise Similarity} chart to depict representation changes across the network.

The \emph{Class-wise Accuracy} chart visualizes each model’s per-class accuracy difference using a diverging bar chart (\figref{metrics}A). 
A dotted line denotes the average accuracy difference across all retain classes, enabling a quick assessment of whether the forget class is adequately forgotten without overly compromising the retain classes. 
As shown in \figref{metrics}A, \textcolor{green}{Model~A} achieves higher accuracy on the ``airplane'' (forget class) and ``dog'' (one of the retain classes), while \textcolor{purple}{Model~B} outperforms it on most other retain classes. 
Hovering over a bar reveals the exact per-class accuracies and differences for both models. 
On the right, results on the test dataset are juxtaposed, helping users simultaneously gauge overall generalization. 
This design highlights the high-level trade-off between under-forgetting (insufficient removal) and over-forgetting (excessive removal).

\begin{figure}[t]
    \centering
    \includegraphics[width=\linewidth]{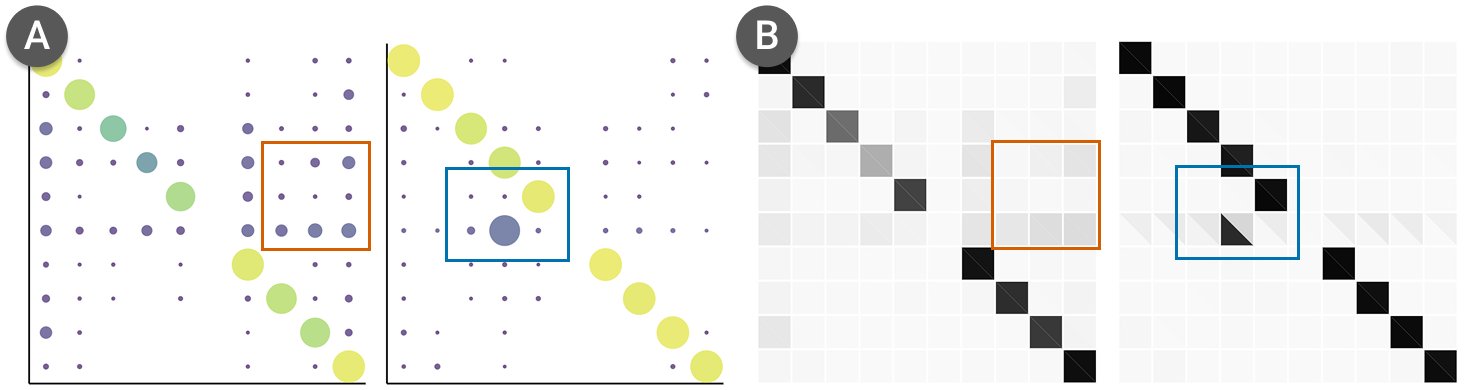}
    \caption{Initial and final designs of the \emph{Prediction Matrix}, both encoding predicted proportion and average confidence. 
    The initial design (A) uses circle size and color, while the final design (B) arranges them diagonally.}
    \label{fig:matrix}
    \vspace{-2mm}
\end{figure}

The \emph{Prediction Matrix} provides a detailed view of how a model assigns predicted classes for each true class (\figref{metrics}B). 
Traditionally, confusion matrices encode the true class on the row axis and the predicted class on the column axis, using color to represent the proportion of samples falling into each cell. 
However, we aim to incorporate not just these proportions but also the model’s confidence—merely observing classification errors does not reveal how confidently a model is mistaken. 
This finer observation relates to a commonly recognized issue of confidence calibration~\cite{guo2017calibration}, which can be exacerbated by the distribution shifts inherent to the unlearning process. As shown in \figref{metrics}B, users can compare two models' prediction patterns side-by-side; for instance, in the tooltip of \textcolor{purple}{Model~B}, one cell has a predicted proportion of 0.560 yet an average confidence of only 0.160, unlike \textcolor{green}{Model~A}, illustrating how large a mismatch can remain hidden if only proportions are considered.

Our initial design took inspiration from prior work~\cite{hohman2020understanding, wangDodrioExploringTransformer2021} that uses size and color for a double-encoding heatmap, representing predicted proportions with circle size and average confidence with color (\figref{matrix}A). 
However, during our study, we identified a perception issue: when both visual channels encoded information within a single glyph, users struggled to discern subtle variations in confidence (color), particularly when the proportions (size) were small, a common scenario when unlearning scatters predictions across multiple classes.
Therefore, we chose to use the diagonal-split encoding~\cite{alper2013weighted,zhao2015matrixwave}, which addresses this by spatially separating the two channels (\figref{matrix}B): predicted proportion in the lower-left and average confidence in the upper-right. This design preserves color detail more clearly (\figref{matrix}-Orange Box) and makes their relationship more apparent (\figref{matrix}-Blue Box).

The \emph{Layer-wise Similarity} chart illuminates changes in internal representations across the network, offering a perspective distinct from previous output-level views (\figref{metrics}C). For a set of representative layers (e.g., the initial, final, and the boundary layers of each residual block in ResNet-18), we measure the Centered Kernel Alignment (CKA)~\cite{kornblith2019similarity} of each selected model against two references: the original model, to gauge the magnitude of representational change, and the retrained model, to assess alignment with ideal representations. 
We chose to use CKA instead of measuring the distances between parameters~\cite{bernstein2020distance}, as CKA's permutation-invariant comparison of activation spaces can facilitate the comparison of functionally similar models.

This chart allows users to visually track at the layer level when and how the representations for the forget and retain classes diverge. For example, in \figref{metrics}C, both models exhibit similar representational changes in the final fully connected layer. However, an internal layer (\texttt{layer4.0}) shows a substantial drop in CKA for \textcolor{green}{Model~A} (0.246) but not for \textcolor{purple}{Model~B} (0.822). This indicates that \textcolor{purple}{Model~B} continues to encode much of the forget class’s features in this layer, while \textcolor{green}{Model~A} removes them more substantially.

\begin{figure*}[t]
\centering
\includegraphics[width=\textwidth]{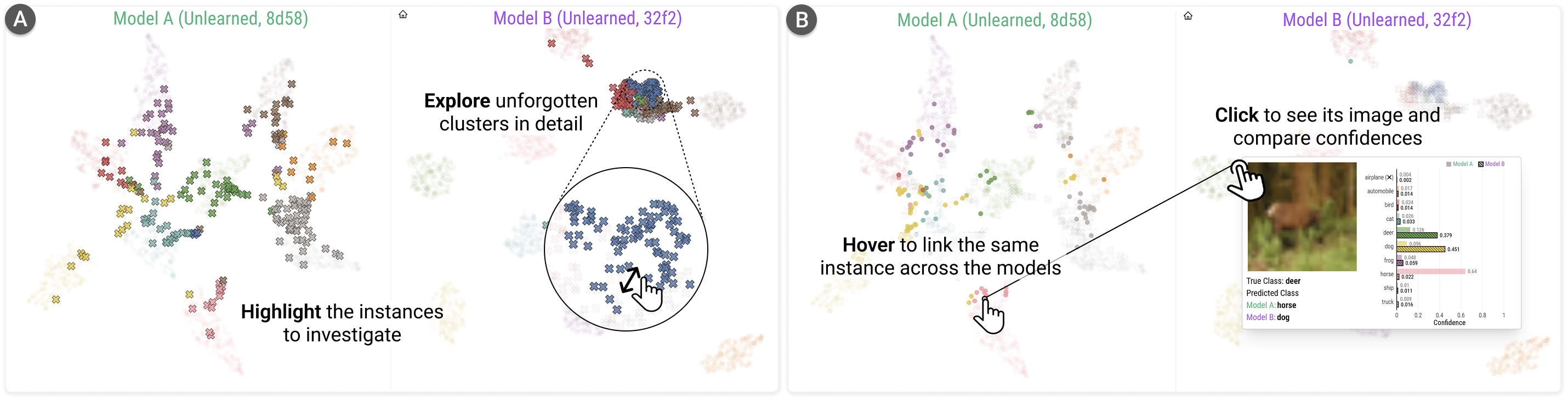}
\caption{
Comparative visual analysis of feature space for two unlearned models in the \textbf{Embedding Space} view. 
(A) Highlighting the forget class instances reveals how their feature distribution differs between them. For instance, \textcolor{purple}{Model~B} (right) contains a dense, unforgotten cluster that can be explored in detail. 
(B) Linking interactions enable direct comparison. Hovering connects the same instance across models, while clicking on an instance reveals its image and compares predicted confidences.
}
\label{fig:embedding}
\vspace{-2mm}
\end{figure*}

The \textbf{Embedding Space} view visualizes the feature space to analyze and compare each model’s decision boundary  (\figref{embedding}). 
We pass each data sample through \textcolor{green}{Model~A} and \textcolor{purple}{Model~B} up to the penultimate layer to obtain high-dimensional feature vectors, then reduce them to two dimensions and display the resulting scatterplots side-by-side. 
%for direct comparison.
% Because final predictions rely on a linear transformation of these penultimate layer outputs, visualizing them clarifies how the decision boundary is formed.
% Placing the two scatterplots this way lets users see how an unlearned model’s feature vectors shift from the original model and determine whether they resemble the retrained model’s decision boundary.
Because final predictions are a linear transformation of these penultimate representations, this view reveals how an unlearned model’s features shift from the original model and whether they align with the retrained model.
\rev{For example, after successful unlearning, the forget class embeddings should align with the retrained model, whereas persistent forget class clusters or distorted retain class clusters (e.g., density changes) can indicate under- or over-forgetting.}
We use UMAP~\cite{mcinnes2018umap} for dimensionality reduction because it is known to preserve global structure more effectively than other techniques, such as $t$-SNE~\cite{van2008visualizing}, aiding in locating forget class samples that might scatter unpredictably.

In this view, all training instances are displayed by default (\figref{teaser}D), but users can activate a highlight mode to focus on specific subsets (\figref{embedding}A): \emph{Target to Forget} (all forget class samples), \emph{Successfully Forgotten} (forget class samples predicted differently from the true label), \emph{Not Forgotten} (forget class samples still predicted as their true label), or \emph{Overly Forgotten} (retain class samples misclassified). 
Users can explore shifts in the feature space to identify areas of interest, such as dense unforgotten clusters (\figref{embedding}A-right), and perform instance-level comparisons by linking corresponding instances to track a single sample’s changing predictions and confidences between the two models (\figref{embedding}B).
% \rev{To keep the view focused on sample-level analysis, we adopt this design while noting that for much larger datasets the view could be complemented with embedding-based summaries (e.g., aggregated embeddings~\cite{hohman2020understanding} or concept clustering~\cite{huang2022conceptexplainer}).}
%\rev{We adopt this design to support sample-specific diagnosis; for much larger datasets, the view could be complemented with embedding-based summaries (e.g., aggregated embeddings~\cite{hohman2020understanding} or concept clustering~\cite{huang2022conceptexplainer}).}
%Users can explore shifts in the feature space to identify areas of interest, such as dense unforgotten clusters (\figref{embedding}A), and perform instance-level comparisons by linking corresponding instances to track a single sample’s changing predictions and confidences between the two models (\figref{embedding}B).

\subsubsection{Attack Simulation (the \textit{Attack} stage, \taskref{T5})}

\begin{figure*}[t]
    \centering
    \includegraphics[width=\textwidth]{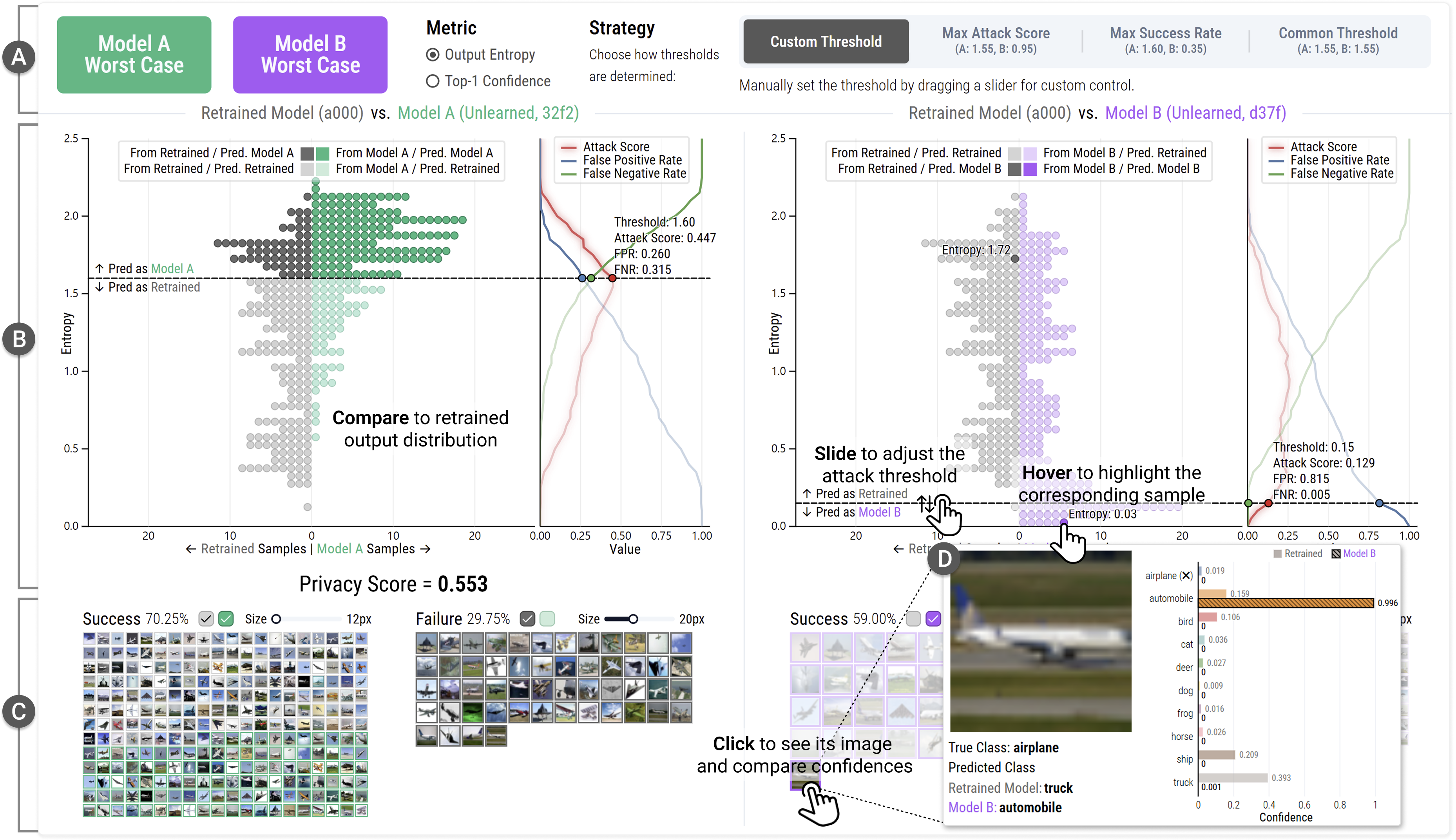}
    \caption{
        The \textbf{Attack Simulation} view to assess two unlearned models against the retrained model via MIAs.
        Gray, \textcolor{green}{green}, and \textcolor{purple}{purple} dots indicate samples output by the retrained model, \textcolor{green}{Model~A}, and \textcolor{purple}{Model~B}, respectively.
        (A) Users configure the attack metric and threshold-setting strategy, and can use a button to inspect the worst-case scenario.
        (B) Dot plots display each model’s output distribution compared with the retrained model, with FPR/FNR and attack score tracked by a threshold slider.
        (C) ``Success'' and ``Failure'' samples appear in a grid for instance-level inspection.
        (D) Clicking on any sample reveals its original image and compares the retrained and unlearned models’ predicted confidence.
    }
     
    \label{fig:attack}
    \vspace{-2mm}
\end{figure*}

While previous views focused on comparing the two selected models from multiple perspectives, the \textbf{Attack Simulation} view compares their output distributions against the retrained model and applies MIAs to discriminate between them (\figref{attack}). 
We draw inspiration from Wattenberg et al.~\cite{wattenberg2016attacking} for the interface design.

The procedure for defining per-threshold attack scores and a privacy score is detailed in \secref{PS}. 
Accordingly, we obtain each forget class sample’s logit vector and transform it into confidence or entropy value (\equref{confidence}, \equref{entropy}). 
Each transformed sample appears as a dot in two plots: (1) retrained model vs.\ \textcolor{green}{Model A} and (2) retrained model vs.\ \textcolor{purple}{Model B}. 
A threshold slider overlays each plot, and the corresponding FPR, FNR, and attack score (\equref{epsilon}, \equref{AS-FQS-t}) are shown on the right as line charts, providing a metric view of the attack (\figref{attack}B).

Users can perform various attacks in this view (\figref{attack}A).
They may select whether to use entropy or confidence and choose different threshold-setting strategies.
For example, they can set a threshold to maximize the attack score based on our defined FPR and FNR, select a threshold with higher overall success, or apply a common threshold for both \textcolor{green}{Model A} and \textcolor{purple}{Model B} \rev{yielding the highest attack score to compare their vulnerability under the same threshold setting.}
%to achieve the highest attack score under identical conditions.
Ultimately, across two attack metrics, threshold directions, and strategies, the worst-case scenario—where the model is most vulnerable—defines the \emph{WCPS}, \rev{and the user can automatically apply the configuration via a dedicated button.}
% and the user can check exactly which configuration leads to that worst case via a dedicated button.

To move beyond distribution-level inspection, we provide interaction and a grid view for instance-level details (\figref{attack}C). Attack success or failure is shown through small image thumbnails; hovering over an image or dot highlights the corresponding instance in the other model, letting users view precise numeric values and compare each sample’s position within the distribution. Clicking a sample displays its original image and compares the retrained and unlearned models’ per-class confidence scores (\figref{attack}D).

As an example, \figref{attack}B shows a scenario on CIFAR-10 where ``airplane'' is the forget class. The retrained model exhibits a diverse range of entropy values for those samples. By contrast, \textcolor{green}{Model~A} often produces higher entropy for the forget class, making the two distributions more easily distinguishable. Meanwhile, \textcolor{purple}{Model~B} appears broadly similar to the retrained model yet features a cluster of low-entropy outliers. By adjusting the threshold, users can isolate these outliers in the grid, apply filters, and inspect them individually. In \figref{attack}D, for instance, \textcolor{purple}{Model~B} assigns strong confidence to ``automobile,'' whereas the retrained model spreads its confidence among multiple classes. Under most thresholds, this distinct difference makes the sample easily singled out, indicating a higher privacy risk.

% \vspace{10mm}
\subsection{Implementation Details}
% \del{\footnote{\url{https://react.dev}}}
% \del{\footnote{\url{https://react.dev}}}
% \del{\footnote{\url{https://ui.shadcn.com}}}
% \del{\footnote{\url{https://fastapi.tiangolo.com}}}
Our implementation consists of a React.js frontend that provides visualizations powered by D3.js and follows the Shadcn/UI design system, and a FastAPI backend that manages model evaluation, unlearning operations, and data handling.
The system provides pre-computed results for system-provided models stored as JSON files, with their checkpoints automatically downloaded from Hugging Face repositories\footnote{\url{https://huggingface.co/jaeunglee/resnet18-cifar10-unlearning}, \url{https://huggingface.co/Yurim0507/resnet18-fashionmnist-unlearning}, \url{https://huggingface.co/Yurim0507/vit-base-16-cifar10-unlearning}} upon backend initialization. 
When users configure unlearning parameters and initiate execution through the UI, the frontend sends API calls to trigger real-time unlearning on the backend.
\rev{On a single NVIDIA A6000 GPU, unlearning typically takes from several seconds to about 5 minutes depending on the method and architecture. The subsequent evaluation pipeline (e.g., CKA computation, dimensionality reduction) requires approximately 30 seconds.}
Upon completion, the backend saves both model checkpoints and metrics for visualization as JSON files. The JSON is then transmitted to the frontend for rendering.
Additionally, users can add custom MU methods by implementing Python hooks with minimal boilerplate code; once registered, these methods appear in the UI alongside existing MU methods. The source code is publicly available at \url{https://github.com/gnueaj/Machine-Unlearning-Comparator}.
\section{Case Study}
To evaluate if and how our system assists users in understanding the behavior of MU methods, we conducted a case study in collaboration with two experts introduced in \secref{domain}. 
To the best of our knowledge, this study presents the first comparative visual analysis of prominent MU methods to enhance their transparency and interpretability~\cite{shaik2025exmu}.
The case study consists of two stages. 
In the first \textit{analysis} stage, the experts freely used our system to compare the behavior of five MU methods: three baseline methods (FT, RL, and GA) and two state-of-the-art MU methods (SCRUB and SalUn) as described in \secref{method}.
To examine whether the insights from the analysis stage could lead to actual improvements in MU performance, we conducted a second, \textit{improvement} stage, in which the experts developed a novel MU method based on their findings from the analysis stage. For the case study, we employed a ResNet-18 architecture on the CIFAR-10 dataset, designating various classes (e.g., ``airplane,'' ``automobile,'' or ``deer'') as the forget class one at a time. 

\subsection{Analysis Stage: Comparing the Existing MU Methods}
In the analysis stage, the two experts collaboratively used our system for a week to (1) investigate the behavior of each MU method and (2) evaluate them based on MU principles.
We summarize the visualizations the experts employed (\figref{finding1}, \fignref{finding2}, \fignref{finding3}, \fignref{finding5}, and \fignref{finding6}) and their corresponding findings (\findref{1}–\findnref{6}) below. 
These findings later informed the development of a novel MU method.

\subsubsection{Retrained Model Behavior (Finding 1)}
\label{finding:1}
\begin{figure}[t]
    \centering
    \includegraphics[width=\linewidth]{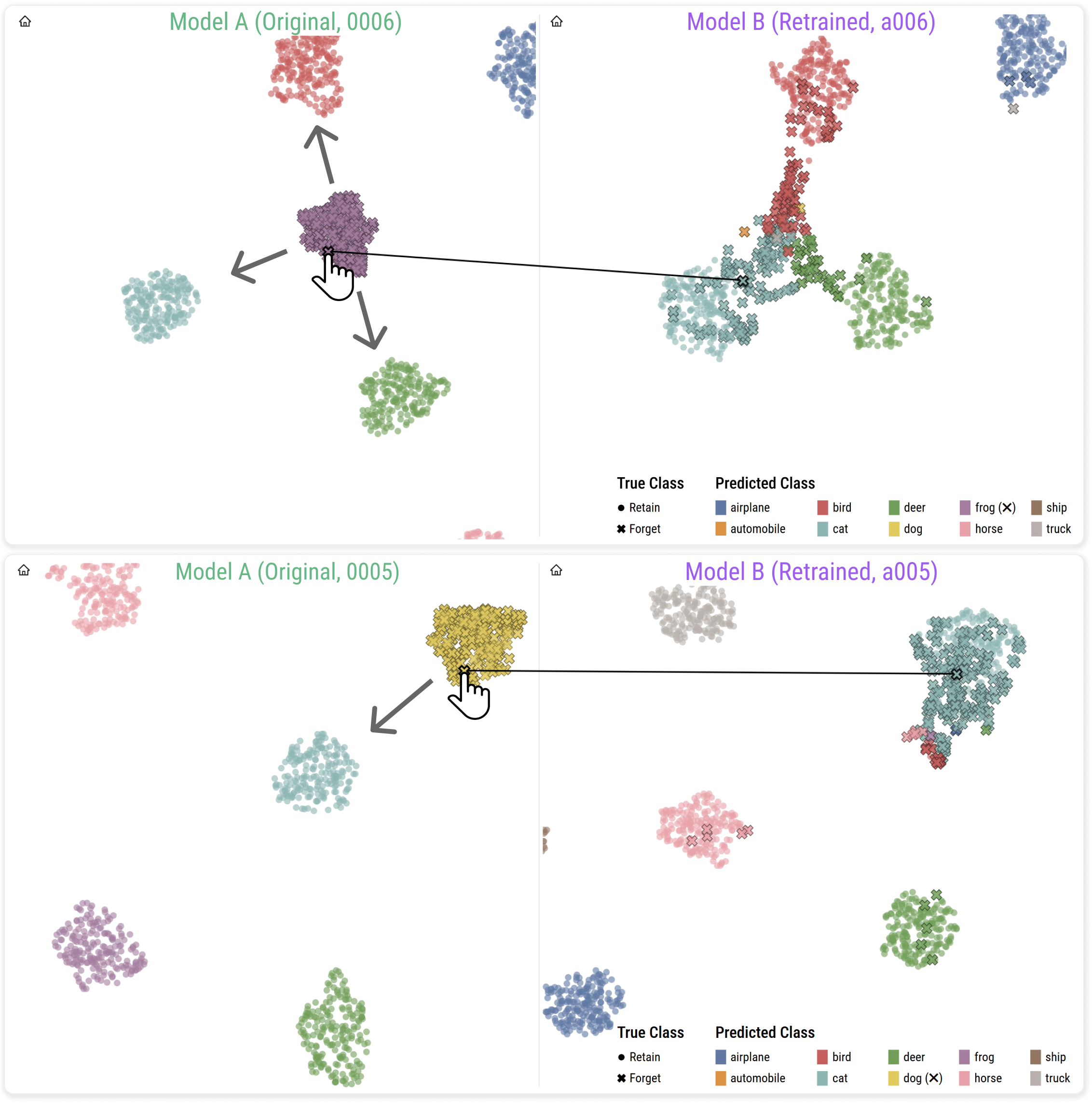}
    \caption{
    \emph{Finding 1} --
    Analyzing embedding shifts between the original model (\textcolor{green}{Model~A}) and the retrained model (\textcolor{purple}{Model~B}).
    \textbf{Top}: When ``frog'' is the forget class, embeddings from this class are redistributed into nearby clusters (e.g., bird, cat, deer).
    \textbf{Bottom}: When ``dog'' is the forget class, most embeddings in the retrained model shift toward the cat cluster.
    }
    \label{fig:finding1}
\end{figure}

Before analyzing each MU method in depth, they first compared the retrained model with the original model (\taskref{T2.1}) to understand how the model behaves when it has never seen the forget class.
In one such analysis, as illustrated in \figref{finding1}, the \textbf{Embedding Space} view shows that many of the original model’s embeddings shift toward nearby clusters in the retrained model.
This comparison provides a reference point for evaluating how MU methods remove the forget class.

\subsubsection{Hyperparameter and Class-wise Trade-offs (Finding 2)}
\label{finding:2}
\begin{figure}[t]
    \centering
    \includegraphics[width=\linewidth]{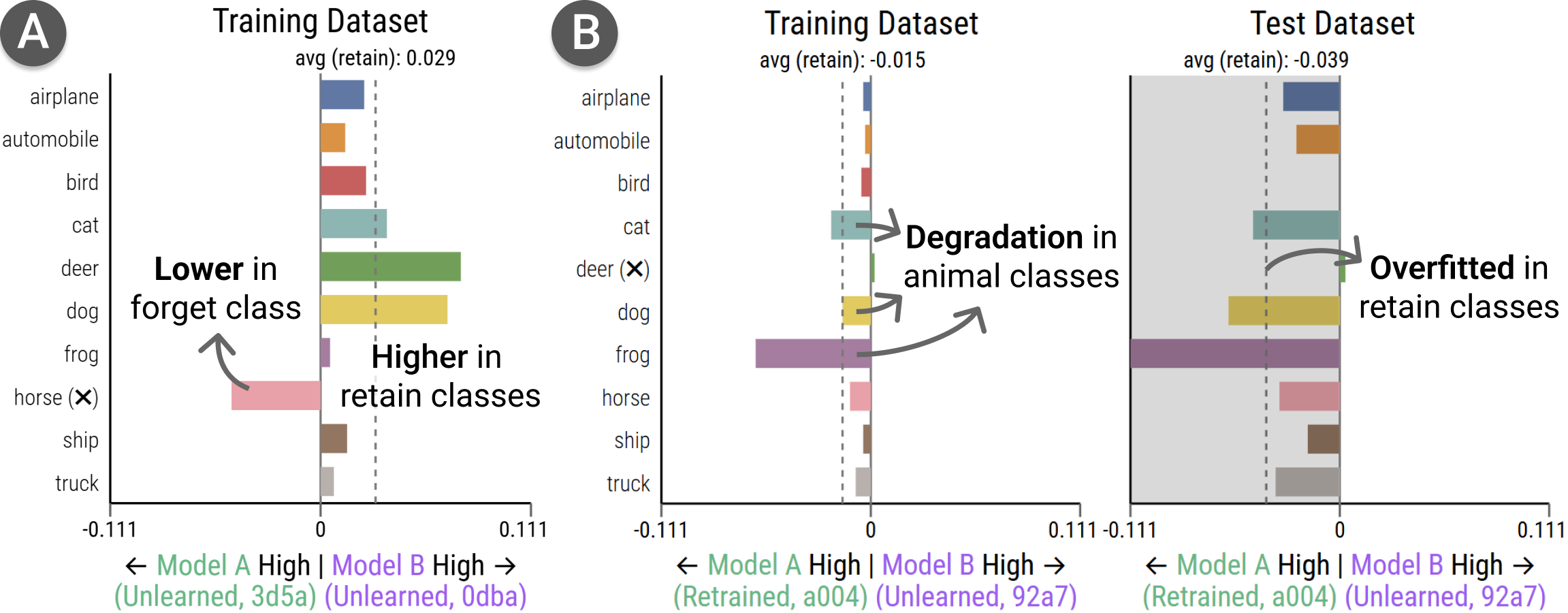}
    \caption{
    \emph{Finding 2} -- Examining the \emph{Class-wise Accuracy} trade-offs.
(A) Two GA variants with different batch sizes: compared to \textcolor{green}{Model~A} (small batch), \textcolor{purple}{Model~B} (large batch) exhibits lower accuracy on the forget class with higher accuracy on the retain classes, indicating better unlearning.
(B) An unlearned model, \textcolor{purple}{Model~B} (FT), is compared to \textcolor{green}{Model~A} (Retrain) in both training and test datasets. It reveals performance degradation on semantically similar classes to the forget class (B-left) and overfitting in the retain classes (B-right).}
    \label{fig:finding2}
    \vspace{-2mm}
\end{figure}

\begin{figure}[t]
  \centering
  \includegraphics[width=0.99\linewidth]{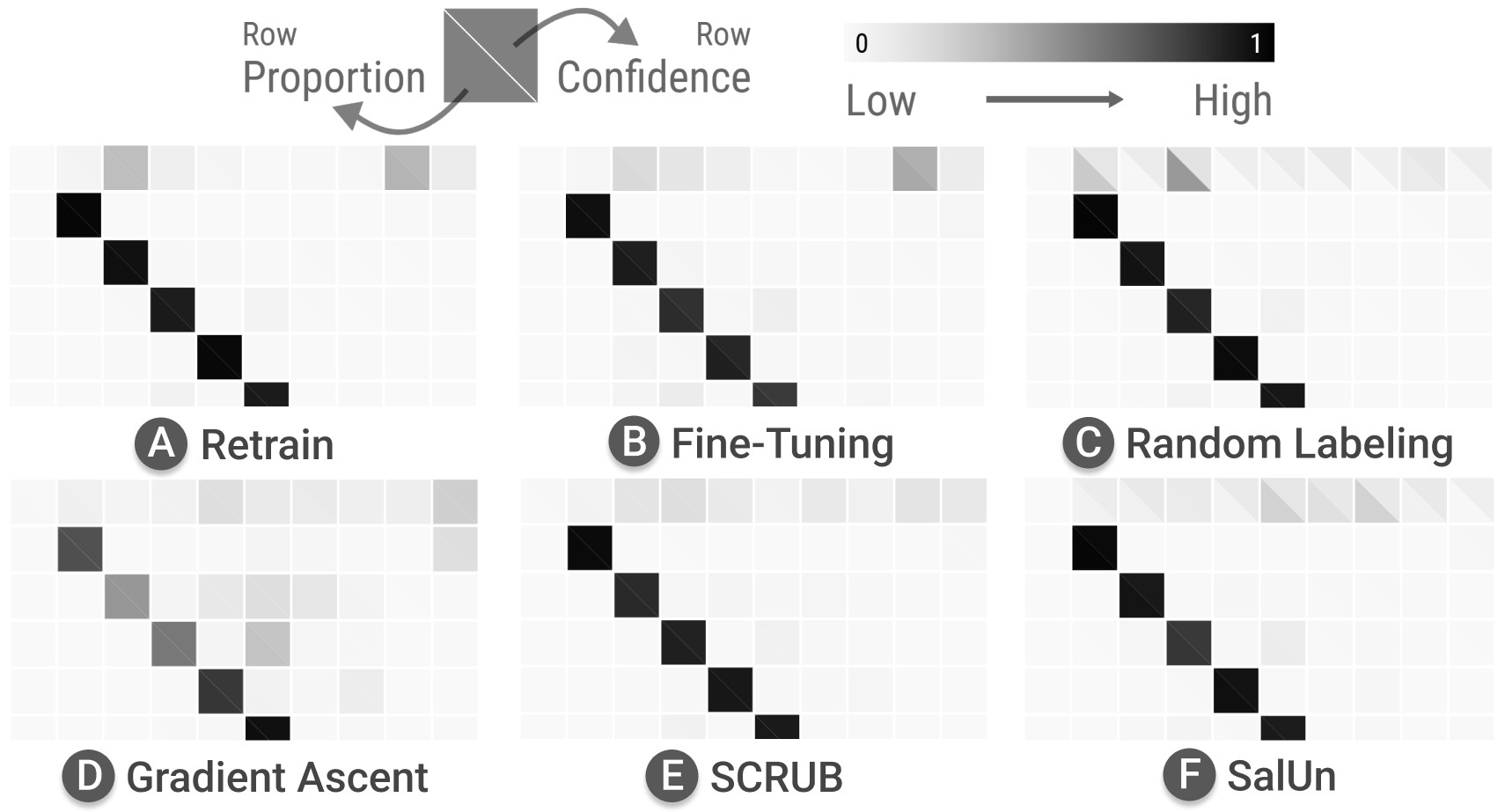}
  \caption{
  \emph{Finding 3} --
  Investigating misclassification patterns and detecting confidence mismatches.
  The first row is the forget class ``airplane,'' and each column represents the predicted class.
  By comparing this entire row to the retrained model (A), users can see how similarly each unlearned method allocates the forget class.
  Notably, brightness contrast between the two triangles in a single cell reveals inconsistencies between confidence and predicted proportion (C and F).
  }
  \label{fig:finding3}
\end{figure}
They tested different hyperparameter combinations for each method using the \textbf{Model Builder} and the \textbf{Model Screening} view, examining class-specific impacts through the \emph{Class-wise Accuracy} chart (\taskref{T1}, \taskref{T2.3}).
During this process, they observed that different MU methods performed best under distinct hyperparameter settings.
For example, GA performed well with a larger batch size, higher learning rate, and fewer training epochs (\figref{finding2}A).
This may be due to larger batches providing more stable gradients, enabling targeted forgetting of forget class representations.
In contrast, FT showed better results with more training epochs, though this also made it more prone to overfitting (\figref{finding2}B-right).
They also found that retain classes sharing similar features with the forget class (e.g., ``deer'' vs. animals such as ``frog'' or ``dog'') exhibited greater accuracy degradation (\figref{finding2}B-left). These results highlight how the system helps users refine hyperparameters and better understand the inherent class-wise trade-offs of each method.

\subsubsection{Misclassification Patterns (Finding 3)}
\label{finding:3}
After tuning the hyperparameters for each MU method, they became particularly interested in how confident each method was in its predictions for the forget class versus the retain classes (\taskref{T3}). To investigate this, they used the \emph{Prediction Matrix} (\figref{finding3}).
They found that most methods produced stable outputs (i.e., proportions closely matched their corresponding confidence) for the retain classes.
However, the methods showed distinct differences in the forget class.
While FT, GA, and SCRUB (\figref{finding3}B, D, and E) all yielded stable outputs, their prediction patterns diverged: FT and SCRUB were similar to the retrained model (\figref{finding3}A), whereas GA's were not.
Notably, RL and SalUn (\figref{finding3}C and F), both relying on random labeling, exhibited a clear mismatch between predicted proportions and confidence for the forget class (i.e., the brightness contrast between the two triangles in the first-row cells).
This mismatch highlights an issue of poor confidence calibration~\cite{guo2017calibration}: \rev{high prediction proportions can mask low confidence, so these signals should be treated as diagnostic indicators and validated or calibrated} before informing high-stakes decisions such as medical diagnosis.
% such a model appears decisive due to high prediction proportions, while its low confidence values indicate internal uncertainty, leading to unreliable decisions in high-stakes domains such as medical diagnosis.

\subsubsection{Feature Space Shifts (Finding~4)}
\label{finding:4}
\begin{figure}[t]
  \centering
  \includegraphics[width=0.98\linewidth]{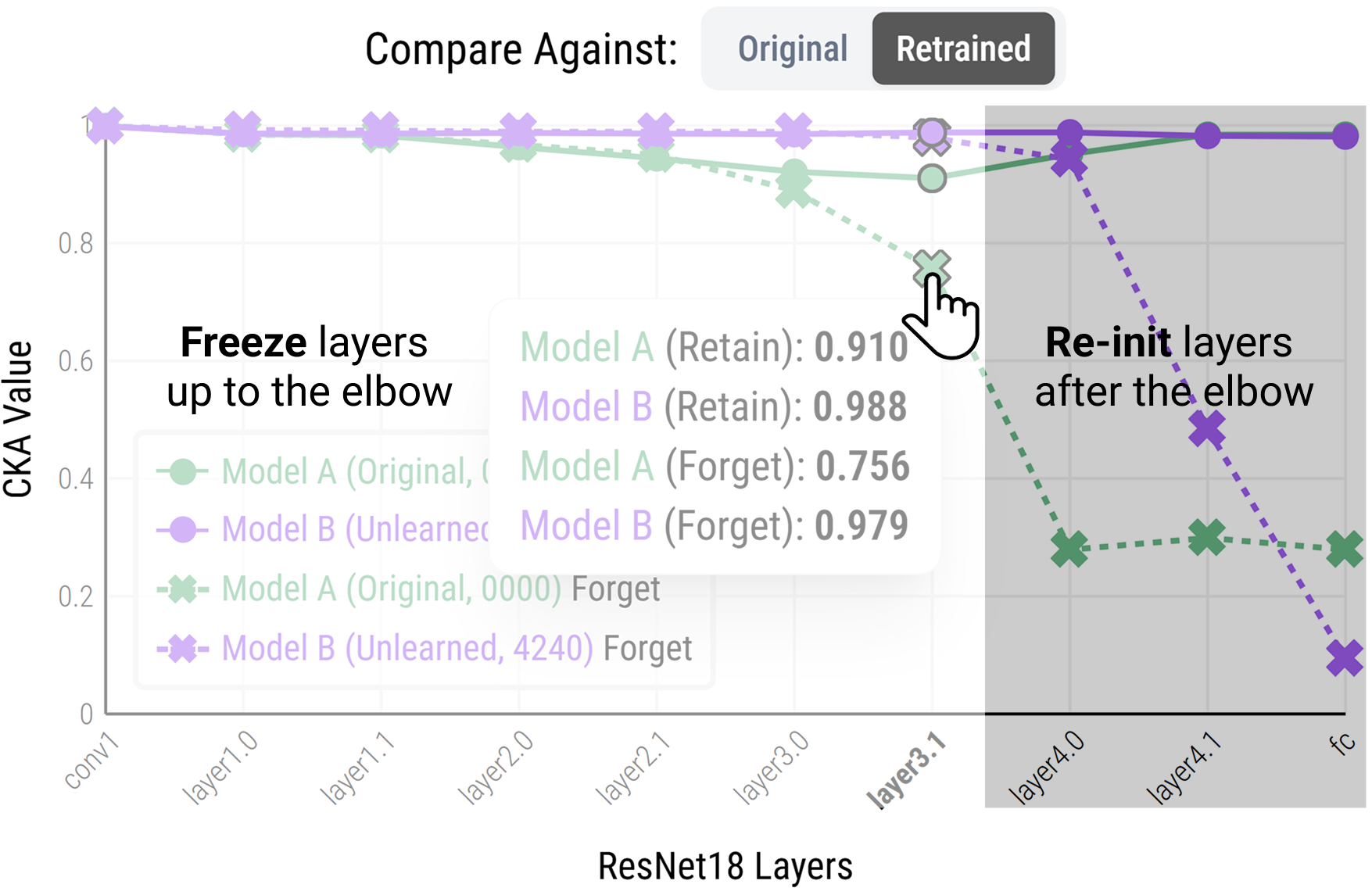}
  \caption{
  \emph{Finding 5} --  Identifying the \emph{Elbow Layer} via Layer-wise similarity analysis. 
  The chart compares similarity of two models against the retrained model: \textcolor{green}{Model~A} (Original) and \textcolor{purple}{Model~B} (RL). The \emph{Elbow Layer} (e.g., \texttt{layer3.1}) is where retain class CKA is minimized, just before the sharp divergence for the forget class. This provides the rationale for our strategy of re-initializing these later, output-centric layers for efficient unlearning.
}

  \label{fig:finding5}
\end{figure}
\begin{figure}[t]
  \centering
  \includegraphics[width=0.99\linewidth]{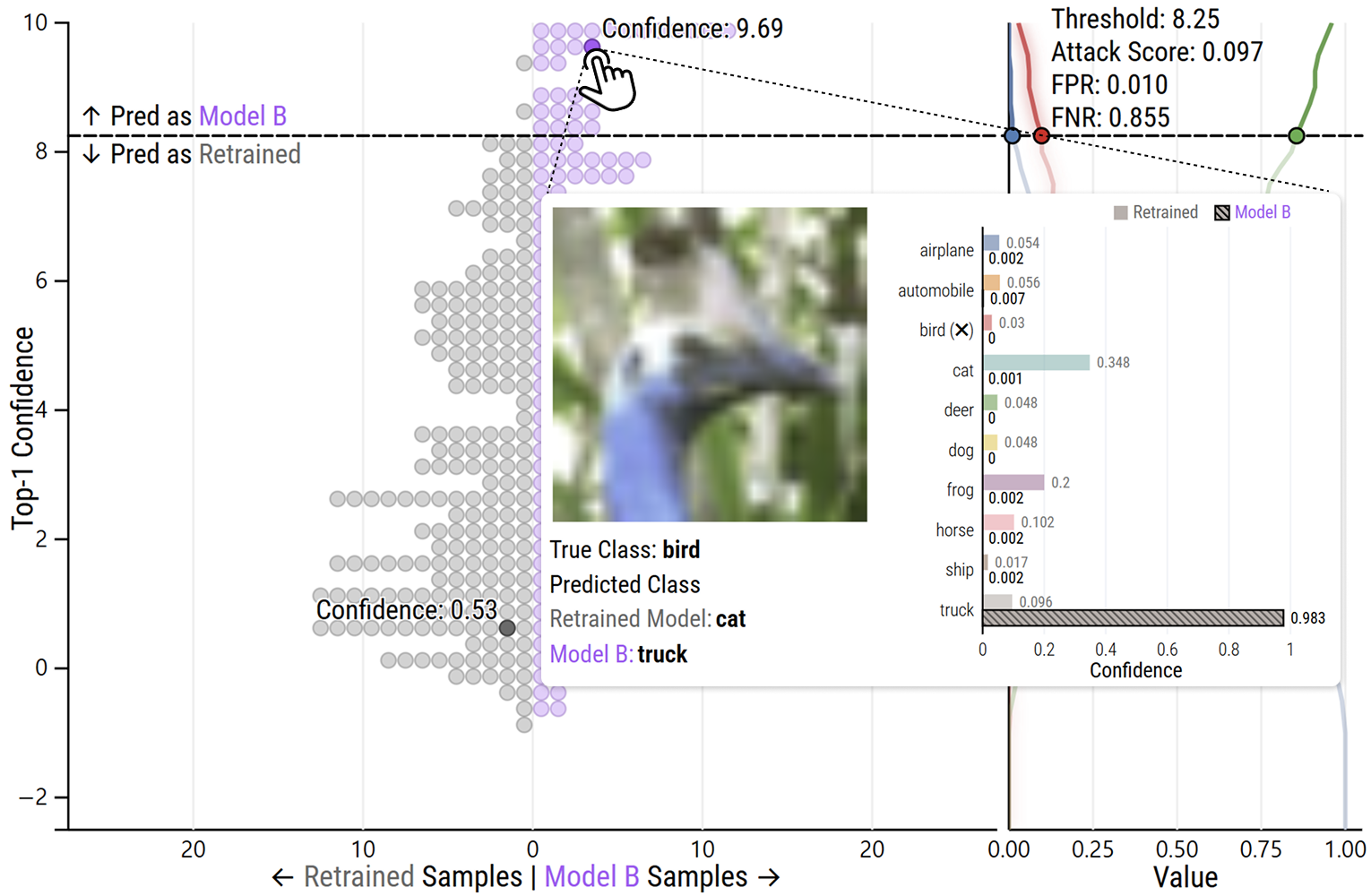}
  \caption{
  \emph{Finding 6} --
   The \textbf{Attack Simulation} view after unlearning of ``bird.'' 
   Samples near the cursor (with top-1 confidence greater than 0.8) exhibit abnormally high confidence for a non-animal class—an outcome rarely observed in predictions from the retrained model. This suggests a potential vulnerability for MIAs.
}
  \label{fig:finding6}
\end{figure}
To understand how each method reshapes the feature space, they examined instance-level embeddings through the \textbf{Embedding Space} view (\taskref{T4}). 
They observed that FT and SCRUB generally shifted embeddings in ways similar to the retrained model (\taskref{T2.2}). However, other methods exhibited distinct behaviors. 
GA did not scatter the forget class embeddings into nearby clusters; instead, they dispersed throughout the entire feature space (\figref{embedding}A-left). 
This suggests that GA erased not only the intrinsic features of the forget class but also unintentionally removed features belonging to nearby samples from the retain classes. 
Consequently, GA appeared to over-forget certain retain classes samples, distributing them irregularly across the space (\figref{embedding}B-left). 
Meanwhile, labeling-based methods such as RL and SalUn maintained high accuracy, but their updates primarily affected the later layers. As seen in \figref{teaser}D, \textcolor{purple}{Model~B} (RL) still formed a distinct cluster for the forget class, indicating that underlying feature representations were largely preserved and only the final classification layers were significantly updated.

\subsubsection{Layer-wise Representation Changes (Finding 5)}
\label{finding:5}
To investigate where these behavioral differences manifest within the network architecture, they further analyzed the \emph{Layer-wise Similarity} chart (\taskref{T4}). 
The analysis confirmed the tendency observed in the \textbf{Embedding Space}; as \figref{metrics}C shows, compared to \textcolor{green}{Model~A} (GA), the changes in \textcolor{purple}{Model~B} (SalUn) were concentrated in the final layers.
Beyond confirming the shallow update pattern, the chart also reveals a transition boundary we term the \emph{Elbow Layer} (\figref{finding5}). By comparing the CKA similarity between the original (\taskref{T2.1}) and retrained models (\taskref{T2.2}), they identified the elbow as the layer where similarity for the retain classes is lowest just before similarity for the forget class begins to diverge sharply (e.g., \texttt{layer3.1} in ResNet-18). This indicates that layers preceding the elbow learn general, foundational features, while subsequent layers specialize in more output-centric features required for class discrimination. This finding offers a principled guideline for setting the value of $k$ in methods that selectively modify a specific number of layers, such as EU-$k$ and CF-$k$~\cite{kurmanji2023towards}. Motivated by the insight, they conducted an experiment: they froze all layers up to the \emph{Elbow Layer}, re-initialized the subsequent layers, and then fine-tuned the model on the retain classes. This strategy resulted in approximately 30\% faster convergence compared to a standard fine-tuning baseline, confirming that the \emph{Elbow Layer} serves as a meaningful boundary for efficient unlearning.

\subsubsection{Privacy Assessment (Finding~6)} 
\label{finding:6}
They employed the \textbf{Attack Simulation} view to assess each method’s privacy beyond just the \emph{WCPS} (\taskref{T5}). In the process, they discovered certain cases where unusually high confidence in the unlearned model made them more vulnerable to MIAs. 
For example, as shown in \figref{finding6}, after unlearning the ``bird'' class using GA, they identified multiple samples whose confidence was relatively low in the retrained model but abnormally high in \textcolor{purple}{Model~B} (the unlearned model), suggesting they could be singled out by an attacker. In a subsequent post-analysis of these high-risk samples, they found that while animal-related features had been removed, certain non-animal features remained, leading the model to misclassify them as trucks with excessively high confidence. This suggests the need for a subsequent step to recover the representations of the samples that were unintentionally damaged during the GA, thereby mitigating the privacy risks posed by these vulnerable misclassifications.

\begin{table*}[t]
  \centering
  % \color{blue} % rev
  \caption{Ablation study of \emph{GU} for two forget classes, ``automobile'' and ``deer''. To reflect practical efficiency in a constrained setting, all variants ran within 3\% of full-retraining time. Each row demonstrates how the steps of \emph{GU}, guided by our visual findings, incrementally enhance the three MU principles. The last row, showing \emph{GU}, achieves strong performance across all principles. It yields a particularly significant improvement in the \emph{WCPS} for the ``automobile'' class, a difficult case where the retrained model consistently misclassifies it as ``truck'' with high confidence.}
  \label{tab:ablation}
  \renewcommand{\arraystretch}{1.1}
  \setlength{\tabcolsep}{5.8pt}
  \begin{tabular}{lccccS[table-format=4.1, detect-weight, mode=text]cccccS[table-format=4.1, detect-weight, mode=text]c}
    \toprule
      & \multicolumn{6}{c}{Forget: automobile} 
        & \multicolumn{6}{c}{Forget: deer}\\
    \cmidrule(lr){2-7}\cmidrule(lr){8-13}
     Method
      & UA$\downarrow$ & RA$\uparrow$ & TUA$\downarrow$ & TRA$\uparrow$ & {RT(s)}$\downarrow$ & WCPS$\uparrow$
      & UA$\downarrow$ & RA$\uparrow$ & TUA$\downarrow$ & TRA$\uparrow$ & {RT(s)}$\downarrow$ & WCPS$\uparrow$ \\
    \midrule
    \rowcolor{black!12}             
     Retrain (Gold Standard)
      & 0.000 & 1.000 & 0.000 & 0.954 & 3524.7 & 1.000  
      & 0.000 & 1.000 & 0.000 & 0.952 & 3513.5 & 1.000 \\
     GA\,+\,FT (Baselines)
      & 0.012 & 0.982 & 0.018 & 0.921 & 104.3 & 0.429  
      & 0.013 & 0.990 & 0.014 & 0.935 & 103.7 & 0.663 \\
     GA\textsubscript{Tuned}\,+\,FT
      & 0.005 & 0.987 & 0.006 & 0.928 &  91.8 & 0.467  
      & 0.005 & 0.995 & 0.004 & 0.942 &  91.8 & 0.737 \\
     Re-Init\,+\,GA\textsubscript{Tuned}\,+\,FT
      & \textbf{0.000} & 0.990 & \textbf{0.000} & 0.933 & \bfseries 71.8 & 0.559     
      & \textbf{0.000} & \textbf{0.995} & \textbf{0.000} & 0.943 & \bfseries 72.3 & 0.751 \\ 
     Re-Init\,+\,GA\textsubscript{Tuned}\,+\,FT\textsubscript{Guided} \emph{(GU)}
      & \textbf{0.000} & \textbf{0.994} & \textbf{0.000} & \textbf{0.938} & 76.6 & \textbf{0.743} 
      & \textbf{0.000} & 0.994 & \textbf{0.000} & \textbf{0.944} & 76.6 & \textbf{0.806} \\
    \bottomrule
  \end{tabular}
\end{table*}

\subsection{Improvement Stage: Developing a Novel MU Method}
Based on the findings from the analysis stage, the experts implemented a novel hybrid MU method, \textit{Guided Unlearning (GU)}.
It consists of three stages: Warm-Up, Forgetting, and Recovery, where the stages respectively perform targeted re-initialization, tuned gradient ascent, and guided fine-tuning. The Warm-Up stage is performed once, while the Forgetting and Recovery stages are alternated for the given epochs.

\textbf{Warm-Up Stage.} \textit{GU} begins by copying the forget class dataset and relabeling its samples using the second-highest logit from the original model. This step aims to approximate the retrained model’s behavior (\findref{1}), while alleviating the mismatch between predicted proportions and confidence that arises from random labeling (\findref{3}). Additionally, motivated by the \emph{Elbow Layer} the experts identified (\findref{5}), the method re-initializes all subsequent high-level layers. This targeted reset would suppress predictions of the forget class,
while earlier layers are preserved to retain essential low-level representations.
A single epoch of fine-tuning is then performed to stabilize the re-initialized layers; this mitigates initial loss spikes in the subsequent Forgetting stage.

\textbf{Forgetting Stage.} To counter the tendency of label-only methods to alter only the final layers (\findref{4}), this stage applies a tuned gradient ascent ($\text{GA}_{\text{Tuned}}$). Informed by \findref{2}, we use a large batch size and learning rate, accumulating all gradients for a single update to minimize side effects on the retain classes.

\textbf{Recovery Stage.} Subsequently, the recovery stage performs a guided fine-tuning ($\text{FT}_{\text{Guided}}$) using a mixture of the relabeled forget class samples from the Warm-Up stage and the samples from the retain classes. By guiding the prediction with relabeled forget samples and recovering representations through fine-tuning on the retain classes, the method mitigates the overconfidence on irrelevant labels (\findref{6}), thereby enhancing privacy and aligning the model's prediction patterns closer to those of a retrained model.
\begin{figure}[t]
  \centering
  \includegraphics[width=\linewidth]{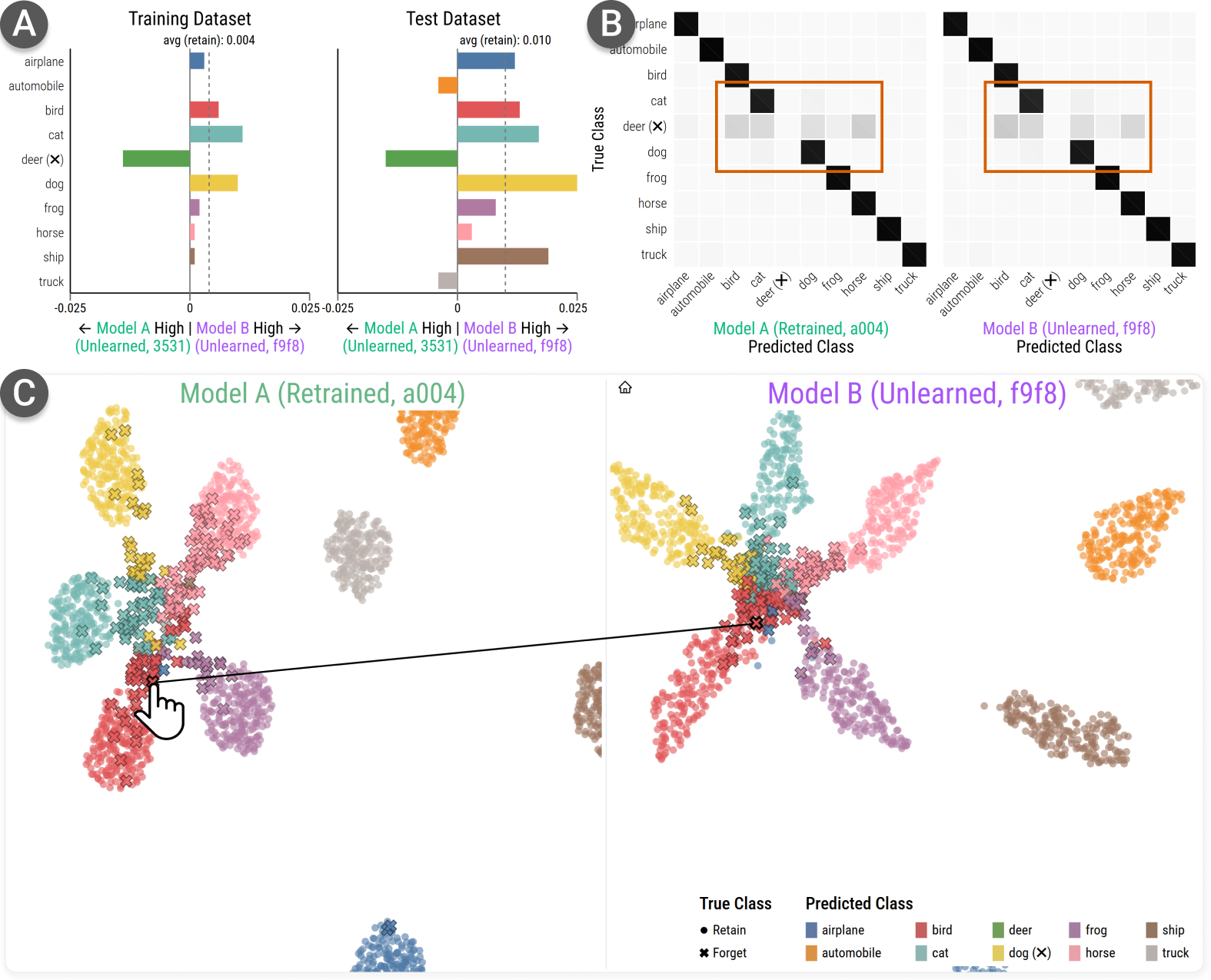}
  \caption{
  Comparison of \emph{GU} with other methods: 
  (A) \emph{GU} (\textcolor{purple}{Model~B}) achieves lower accuracy on the forget class (deer, green bar), and higher accuracy on most retain classes compared to SCRUB (\textcolor{green}{Model~A}).
  (B) \emph{GU} (\textcolor{purple}{Model~B})'s prediction patterns closely resemble the retrained model (\textcolor{green}{Model~A}).
  (C) \emph{GU} (\textcolor{purple}{Model~B})'s feature embeddings exhibit a similar structure to the retrained model (\textcolor{green}{Model~A}).
}
  \label{fig:method}
  \vspace{-2mm}
\end{figure}

We found \textit{GU} outperforms the state-of-the-art method by achieving higher accuracy (\figref{method}A), competitive efficiency, and the highest \emph{WCPS} of 0.913 among all tested methods (surpassing the previous high of 0.876 achieved by SCRUB). Furthermore, pairwise comparisons (\taskref{T2.2}) confirm that \textit{GU} closely matches the retrained model's behavior, specifically its prediction patterns (\figref{method}B), and embeddings (\figref{method}C).
The incremental performance improvements of each step are systematically validated in our ablation study under an efficiency‑constrained setting (\tabref{ablation}).

\section{Expert Feedback}
In addition to the case study, we interviewed four MU researchers who did not participate in the design process (\emph{E1--E4}), with each having over a year of experience in the MU domain. Each interview session lasted 70--90 minutes and included a brief system introduction (20 minutes), a hands-on session (up to 30 minutes), and a concluding discussion on the system’s utility, limitations, and potential research directions (up to 20 minutes).

\textbf{Benefits of the System.}
All researchers acknowledged the need for \emph{Unlearning Comparator} in their research and agreed that its design aligns with the unique characteristics of the domain. 
Specifically, \emph{E2} recognized the importance of pairwise comparisons, stating that \textit{``As machine unlearning research often involves continuous model comparisons, focusing on two methods at a time would be more effective than either examining all methods together or analyzing them one by one.''}
\emph{E3} highlighted the system's alignment with actual research workflows and the utility, remarking that \textit{``It seems to bring researchers close to completing their evaluation pretty quickly, covering roughly 80\% of the necessary steps,''} and particularly highlighting the \textbf{Attack Simulation} view, adding that \textit{``running actual attacks helps us identify vulnerable data points, which could inform more privacy-preserving MU methods.''}
\emph{E4} pointed out the difficulty of managing the variety of metrics in MU evaluation purely in numeric form, noting that \textit{``visualizing these decision factors in a single integrated view would significantly reduce cognitive load.''}

\textbf{Suggestions for Improvement.}
\emph{E1} and \emph{E3}, both working on unlearning for classification models, offered suggestions on extending the system's capabilities.
\emph{E1} explained that it is difficult to monitor whether certain over-forgotten instances are being shifted to different classes, but the \textbf{Embedding Space} interactions \rev{(e.g., linking corresponding instances)} made it simpler to track each sample’s changing predictions. 
Drawing from this, \emph{E1} suggested that visualizing embeddings from lower-level layers might offer useful insights. 
While we currently focus on the penultimate layer for efficient iterative workflow given the computational cost of processing high-dimensional lower-level layers, visualizing these layers could be a valuable option as computational efficiency improves. Also, we note that such 2D projections of high-dimensional data should be interpreted with caution, as projections can sometimes distort the original high-dimensional structure.
In terms of dataset scalability, \emph{E3} noted that the current system is optimized for datasets with ten classes and suggested that supporting more classes would facilitate dataset expansion and tasks like sub-class unlearning.
We acknowledge this scalability challenge as a promising 
future direction.

\section{Limitations and Future Work}
\label{sec:lim}
Additionally, we gathered insights from both the researchers and existing literature regarding our system’s limitations and avenues for future work.

\textbf{Supporting Different MU Tasks.}
Our interface focuses on class-wise unlearning, but researchers have asked about broader scenarios such as multi-class unlearning, instance-wise unlearning, or sub-class unlearning. 
\rev{Specifically, for sub-class unlearning, our \textbf{Prediction Matrix} could be extended by adopting hierarchical confusion matrix designs as in Neo \cite{gortler2022neo} to reveal inter-class misclassifications.}
Extending the interface to handle these would pose new design challenges.

\textbf{Improving Privacy Evaluation.}
Quantifying privacy of MU methods remains challenging. Defining a privacy score and accurately estimating it in practice is still an open problem; this includes the distribution-level indistinguishability approach we adopted. Also, while our system focuses on black-box MIAs, we could consider scenarios where the attacker has access to the model weights (i.e., white-box attacks). 
Such an assumption would enable the evaluation of broader threats including model-inversion attacks~\cite{hu2024learn} or attacks on the unlearning process itself through malicious unlearning requests~\cite{zhao2023static}. 
Integrating these scenarios may refine privacy evaluation and support more comprehensive assessments of MU methods.

\rev{
\textbf{Extending to Foundation Model Unlearning.}
As MU research moves toward foundation models, extending our system will require redesign in several aspects.
For data scalability, much larger datasets may benefit from complementing the instance-level \textbf{Embedding Space} view with embedding-based summaries, such as aggregated embeddings~\cite{hohman2020understanding} or concept clustering~\cite{huang2022conceptexplainer}, to support overview and navigation.
For LLMs, supporting them would necessitate views for tracing unlearning-driven behavioral changes, such as token-level, sequence-oriented inspection~\cite{seq2seqvis} or attention-based behavior tracing~\cite{derose2020attention}, together with system optimizations to maintain interactivity under higher computational cost and latency.
Moreover, our current system provides class-wise retrained models as gold-standard references; for foundation models, supplying such references in advance is often impractical at scale.
As an alternative reference, training a retain-only baseline on MU benchmarks such as TOFU~\cite{maini2024tofu} could serve as a practical gold standard.
Given that foundation model unlearning remains an active area with evolving evaluation conventions (e.g., OpenUnlearning~\cite{dorna2025openunlearning}), incorporating such emerging practices can help align comparative MU analysis with real-world scenarios.
}

% \textbf{Evaluation Without Retrained Models.}
% Our current approach assumes access to a retrained model for comparison. In large-scale tasks such as LLM unlearning~\cite{liu2025rethinking}, however, retraining may be infeasible or prohibitively expensive. 
% To address this concern, we could draw inspiration from the TOFU benchmark~\cite{maini2024tofu}. It asks GPT-4 to generate fake author profiles, fine-tunes a pre-trained LLM on them as the ``original model,'' applies unlearning to produce the ``unlearned model,'' and uses the pre-trained LLM (without fine-tuning) as the ``retrained model.''
% Exploring design spaces that evaluate MU methods without a fully retrained model can be an important direction for future research.

\section{Conclusion}
We present \emph{Unlearning Comparator}, a visual analytics system designed to compare different MU methods, guided by our design study with MU researchers and the tasks identified therein.
Our workflow helps researchers build and screen candidate models, inspect class-, instance-, and layer-level behavior, and perform privacy checks via attack simulation.
Through a case study with MU experts, we conducted a visual analysis of prominent MU methods, deriving findings that informed a novel MU method. This new method demonstrated improved adherence to the MU principles of accuracy, efficiency, and privacy.
In addition, interviews with four MU researchers highlighted that pairwise comparisons reduced cognitive load and helped them detect critical samples more efficiently.
% \newpage
\section*{Acknowledgments}
% 조재민
This work was partly supported by the National Research Foundation of Korea (NRF) grant funded by the Korea government (MSIT) (RS-2023-00221186) and by Institute of Information \& communications Technology Planning \& Evaluation (IITP) grant funded by the Korea government (MSIT) (RS-2019-II190421, Artificial Intelligence Graduate School Program (Sungkyunkwan University)).
% 우사이먼 (변경 완료 1/9)
This work was partly supported by Institute for Information \& communication Technology Planning \& evaluation (IITP) grants funded by the Korean government MSIT: (RS-2022-II220688 and RS-2024-00437849).


% % References (start)
% \section{References Section}
% You can use a bibliography generated by BibTeX as a .bbl file.
%  BibTeX documentation can be easily obtained at:
%  http://mirror.ctan.org/biblio/bibtex/contrib/doc/
%  The IEEEtran BibTeX style support page is:
%  http://www.michaelshell.org/tex/ieeetran/bibtex/
 
%  % argument is your BibTeX string definitions and bibliography database(s)
% %\bibliography{IEEEabrv,../bib/paper}
% %

% You can manually copy in the resultant .bbl file and set second argument of $\backslash${\tt{begin}} to the number of references
%  (used to reserve space for the reference number labels box).
% % References (end)

\bibliographystyle{IEEEtran}
\bibliography{refs}
% Biography (start)
% \newpage
 

% \bf{If you include a photo:}
% \newpage
% \vspace{-5mm}
\begin{IEEEbiography}
[{\includegraphics[width=1in,height=1.25in,clip,keepaspectratio]{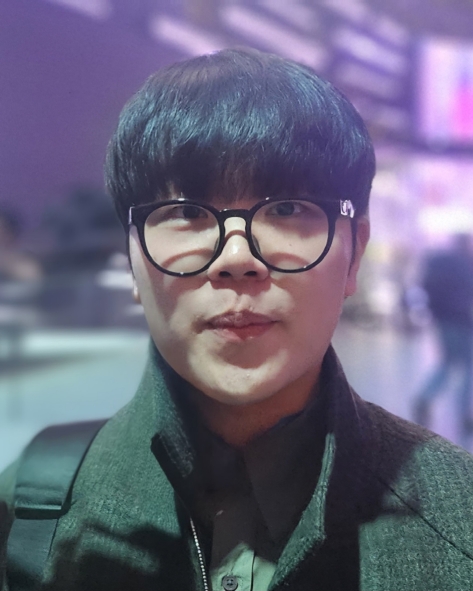}}]
{Jaeung Lee}
is an M.S. student in the Department of Computer Science and Engineering at Sungkyunkwan University. He received his B.S. degree in Computer Science and Engineering from Sungkyunkwan University in 2024. His research interests lie in visualization for machine learning, with a particular focus on how interactive interfaces can support model evaluation and facilitate human-centered decision-making.
\end{IEEEbiography}

\begin{IEEEbiography}
[{\includegraphics[width=1in,height=1.25in,clip,keepaspectratio]{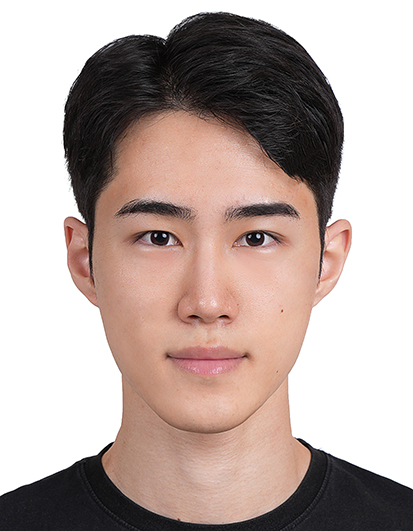}}]
{Suhyeon Yu}
received the B.S. in computer science and engineering from Sungkyunkwan University in 2024. He is currently an M.S. student in the Department of Computer Science at Rice University. His interests include data visualization and cloud computing.
\end{IEEEbiography}

\begin{IEEEbiography}
[{\includegraphics[width=1in,height=1.25in,clip,keepaspectratio]{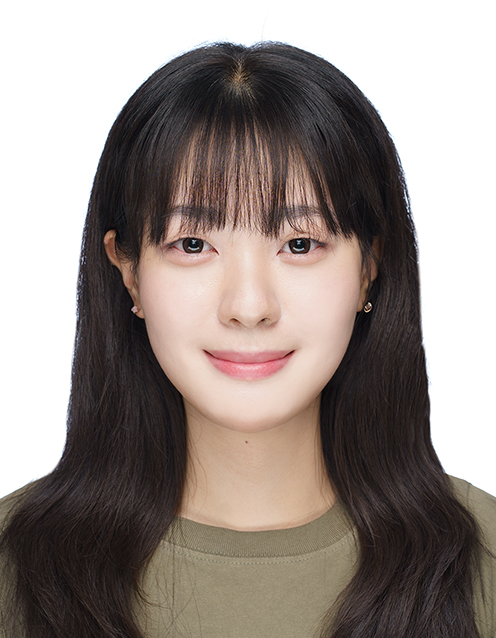}}]
{Yurim Jang}
is an M.S. student in the Department of Artificial Intelligence at Sungkyunkwan University. She received her Bachelor of Science degree in Industrial Engineering from Hongik University in 2024. Her research interests are in machine unlearning and computer vision.
\end{IEEEbiography}

\begin{IEEEbiography}
[{\includegraphics[width=1in,height=1.25in,clip,keepaspectratio]{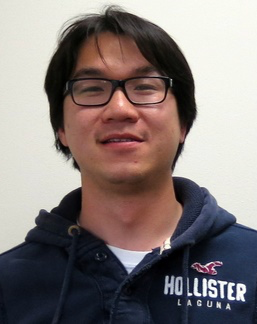}}]
{Simon S. Woo}
is an Associate Professor with the College of Computing and Informatics, Sungkyunkwan University. He received the B.S. degree in electrical engineering from the University of Washington, the M.S. degree in electrical and computer engineering from the University of California, San Diego, and the M.S. and Ph.D. degrees in computer science from the University of Southern California. His research interests include AI security and deepfake detection.
\end{IEEEbiography}

\begin{IEEEbiography}[{\includegraphics [width=1in,height=1.25in,clip, keepaspectratio]{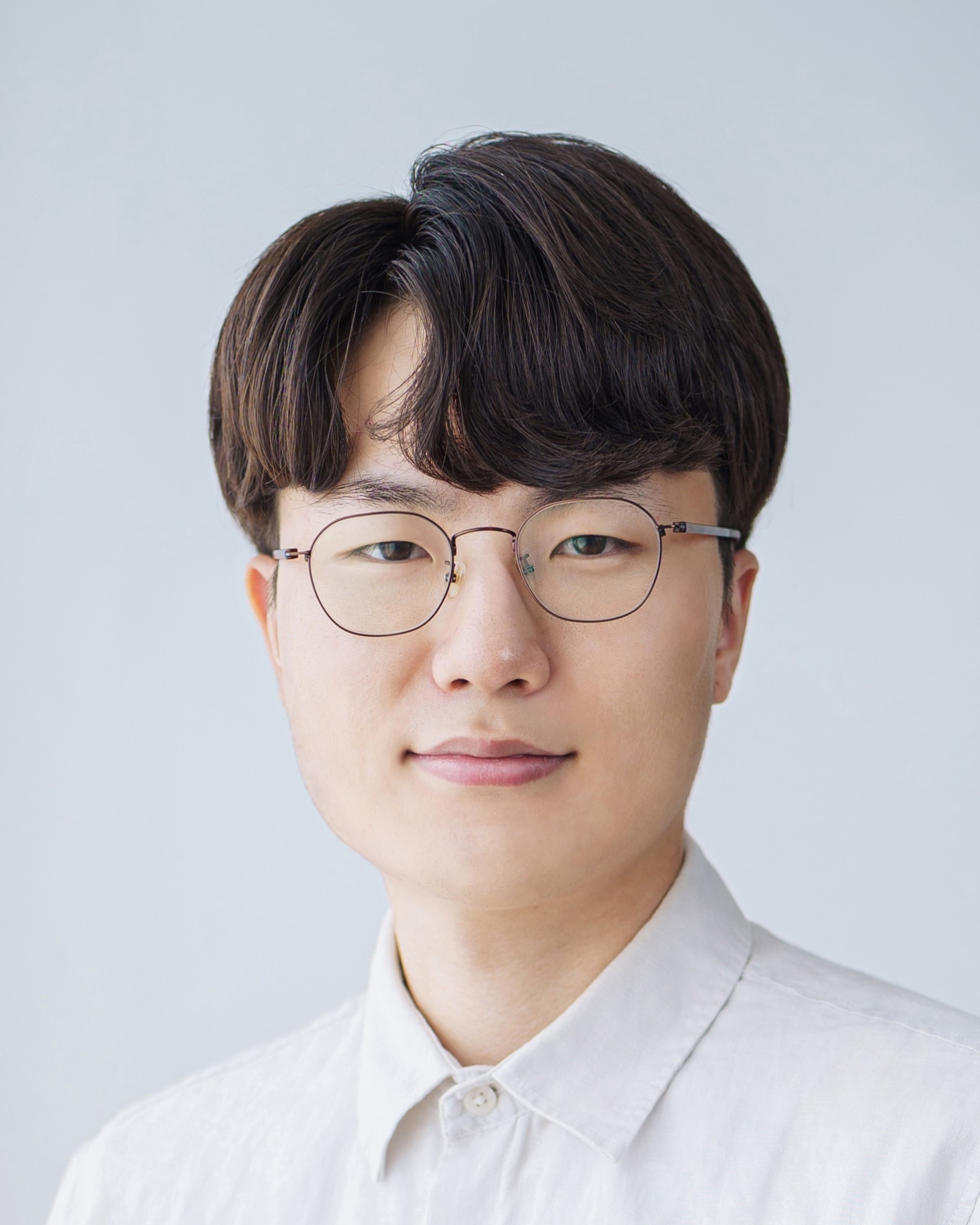}}] {Jaemin Jo} 
received the B.S. and Ph.D. degrees in computer science and engineering from Seoul National University, Seoul, South Korea, in 2014 and 2020, respectively. He is currently an Associate Professor with the College of Computing and Informatics, Sungkyunkwan University, Korea. His research interests include human-computer interaction and large-scale data visualization.
\end{IEEEbiography}


% Biography (end)

\vfill

\end{document}